\documentclass[multphys,vecphys]{svmult}

\usepackage{makeidx}         
\usepackage{graphicx}        
\usepackage{multicol}        
\usepackage[bottom]{footmisc}

\newcommand{\EQ}{\begin{equation}}
\newcommand{\EN}{\end{equation}}
\newcommand{\EQA}{\begin{eqnarray}}
\newcommand{\ENA}{\end{eqnarray}}
\newcommand{\eq}[1]{(\ref{#1})}
\newcommand{\EEq}[1]{Equation~(\ref{#1})}
\newcommand{\Eq}[1]{equation~(\ref{#1})}
\newcommand{\Eqs}[2]{equations~(\ref{#1}) and~(\ref{#2})}

\newcommand{\Sec}[1]{\S\,\ref{#1}}

\newcommand{\Fig}[1]{Figure~\ref{#1}}
\newcommand{\FFig}[1]{Figure~\ref{#1}}
\newcommand{\Tab}[1]{Table~\ref{#1}}

\newcommand{\bra}[1]{\langle #1\rangle}

\newcommand{\meanemfs}{\overline{\cal E} {}}

\newcommand{\meanEMF}{\overline{\mbox{\boldmath ${\mathcal E}$}} {}}
\newcommand{\meanFF}{\overline{\mbox{\boldmath ${\mathcal F}$}} {}}
\newcommand{\meanA}{\overline{A}}
\newcommand{\meanB}{\overline{B}}

\newcommand{\meanU}{\overline{U}}
\newcommand{\meanW}{\overline{W}}
\newcommand{\meanO}{\overline{\Omega}}

\newcommand{\meanT}{\overline{T}}
\newcommand{\means}{\overline{s}}
\newcommand{\meanrho}{\overline{\rho}}
\newcommand{\meanAA}{\overline{\mbox{\boldmath $A$}}}
\newcommand{\meanBB}{\overline{\mbox{\boldmath $B$}}}

\newcommand{\meanJJ}{\overline{\mbox{\boldmath $J$}}}

\newcommand{\meanUU}{\overline{\mbox{\boldmath $U$}}}
\newcommand{\meanWW}{\overline{\mbox{\boldmath $W$}}}

\newcommand{\pp}{\hat{\mbox{\boldmath $\phi$}} {}}

\newcommand{\uu}{\mbox{\boldmath $u$} {}}

\newcommand{\UU}{\mbox{\boldmath $U$} {}}
\newcommand{\bb}{\mbox{\boldmath $b$} {}}

\newcommand{\BB}{\mbox{\boldmath $B$} {}}

\newcommand{\AAA}{\mbox{\boldmath $A$} {}}
\newcommand{\aaa}{\mbox{\boldmath $a$} {}}
\newcommand{\jj}{\mbox{\boldmath $j$} {}}

\newcommand{\JJ}{\mbox{\boldmath $J$} {}}

\newcommand{\EE}{\mbox{\boldmath $E$} {}}
\newcommand{\FF}{\mbox{\boldmath $F$} {}}

\newcommand{\grav}{\mbox{\boldmath $g$} {}}
\newcommand{\nab}{\mbox{\boldmath $\nabla$} {}}
\newcommand{\oo}{\mbox{\boldmath $\omega$} {}}

\newcommand{\kk}{{\bm{k}}}
\newcommand{\ggamma}{\bm{\gamma}}
\newcommand{\ddelta}{\bm{\delta}}
\newcommand{\OOmega}{\bm{\Omega}}

\newcommand{\FFF}{\mbox{\boldmath ${\cal F}$} {}}

\newcommand{\ii}{{\rm i}}

\newcommand{\DD}{{\rm D} {}}
\newcommand{\dd}{{\rm d} {}}
\newcommand{\const}{{\rm const}  {}}

\def\cs{c_{\rm s}}

\def\sigmaSB{\sigma_{\rm SB}}

\def\half{{\textstyle{1\over2}}}

\def\onethird{{\textstyle{1\over3}}}

\def\fourthird{{\textstyle{4\over3}}}

\newcommand{\W}{\,{\rm W}}

\newcommand{\G}{\,{\rm G}}
\newcommand{\Hz}{\,{\rm Hz}}

\newcommand{\nHz}{\,{\rm nHz}}
\newcommand{\kG}{\,{\rm kG}}

\newcommand{\K}{\,{\rm K}}

\newcommand{\g}{\,{\rm g}}
\newcommand{\s}{\,{\rm s}}

\newcommand{\cm}{\,{\rm cm}}
\newcommand{\m}{\,{\rm m}}
\newcommand{\km}{\,{\rm km}}

\newcommand{\kg}{\,{\rm kg}}
\newcommand{\Mm}{\,{\rm Mm}}

\newcommand{\yr}{\,{\rm yr}}

\newcommand{\erg}{\,{\rm erg}}

\newcommand{\J}{\,{\rm J}}

\newcommand{\AU}{\,{\rm AU}}

\newcommand{\yan}[5]{, ``#5,'' {\em Astron.\ Nachr.\ }{\bf #2}, #3-#4 (#1).}

\newcommand{\ypnas}[5]{, ``#5,'' {\em Proc.\ Nat.\ Acad.\ Sci.\ }{\bf #2}, #3-#4 (#1).}
\newcommand{\yana}[5]{, ``#5,'' {\em Astron.\ Astrophys.\ }{\bf #2}, #3-#4 (#1).}

\newcommand{\yssr}[5]{, ``#5,'' {\em Spa.\ Sci.\ Rev.\ }{\bf #2}, #3-#4 (#1).}

\newcommand{\ysph}[5]{, ``#5,'' {\em Solar Phys.\ }{\bf #2}, #3-#4 (#1).}
\newcommand{\ysphS}[5]{, ``#5'' {\em Solar Phys.\ }{\bf #2}, #3-#4 (#1).}
\newcommand{\yjetp}[5]{, ``#5,'' {\em Sov.\ Phys.\ JETP }{\bf #2}, #3-#4 (#1).}

\newcommand{\ynat}[5]{, ``#5,'' {\em Nature }{\bf #2}, #3-#4 (#1).}

\newcommand{\yprl}[5]{, ``#5,'' {\em Phys.\ Rev.\ Letters }{\bf #2}, #3-#4 (#1).}

\newcommand{\yapj}[5]{, ``#5,'' {\em Astrophys.\ J.\ }{\bf #2}, #3-#4 (#1).}
\newcommand{\yapjS}[5]{, ``#5,'' {\em Astrophys.\ J.\ }{\bf #2}, #3-#4 (#1)}

\newcommand{\yaraa}[5]{, ``#5,'' {\em Ann.\ Rev.\ Astron.\ Astrophys.\ }{\bf #2}, #3-#4 (#1).}
\newcommand{\yanar}[5]{, ``#5,'' {\em Astron.\ Astrophys.\ Rev.\ }{\bf #2}, #3-#4 (#1).}

\newcommand{\yrpp}[5]{, ``#5,'' {\em Rep.\ Prog.\ Phys.\ }{\bf #2}, #3-#4 (#1).}
\newcommand{\ypf}[5]{, ``#5,'' {\em Phys.\ Fluids }{\bf #2}, #3-#4 (#1).}

\newcommand{\yjour}[6]{, ``#6,'' {\em #2} {\bf #3}, #4-#5 (#1).}

\newcommand{\ybook}[3]{ {\em #2}.\ #3 (#1).}

\usepackage{natbib}
\usepackage{fancybox}
\usepackage{psboxit}
\usepackage{url}
\usepackage{bm}

\makeindex             

\begin{document}

\title*{Solar Interior -- Radial Structure, Rotation, Solar Activity Cycle}
\author{Axel Brandenburg}
\institute{NORDITA,  Roslagstullsbacken 23,
AlbaNova University Center, 106 91 Stockholm, Sweden
\texttt{brandenb@nordita.dk}}
%
%
\maketitle

\begin{abstract}
Some basic properties of the solar convection zone are considered
and the use of helioseismology as an observational tool to determine its
depth and internal angular velocity is discussed.
Aspects of solar magnetism are described and explained
in the framework of dynamo theory.
The main focus is on mean field theories for the Sun's magnetic field
and its differential rotation.
\end{abstract}

\section{Introduction}

The purpose of this chapter is to discuss the conditions that lead to
the magnetic activity observed at the surface of the Sun.
Only to a first approximation is the Sun steady and spherically symmetric.
A more detailed inspection reveals fully three-dimensional small scale
turbulent motions and magnetic fields together with larger scale
flows and magnetic fields that lack any symmetry.
The cause of the large scale and small scale magnetic fields, as well
as large scale circulation and differential rotation, is believed
to be the turbulent convection which, in turn, is caused by the increased
radiative diffusivity turning much of the radiative energy flux into
convective energy flux.

The magnetic field is driven by a self-excited dynamo mechanism, which
converts part of the kinetic energy into magnetic energy.
As in technical dynamos the term `self-excited' refers to the fact
that part of the electric power generated by induction is also used
to sustain the ambient magnetic field around the moving conductors.
How the conversion of kinetic energy into magnetic energy
works will be discussed in some detail in this chapter.
The kinetic energy responsible for this process can be divided into
(i) small scale irregular turbulent motions (convection) and
(ii) large scale differential rotation and meridional circulation.
It is the anisotropy of the small scale motions that is responsible
for making the rotation nonuniform.
Furthermore, lack of mirror symmetry of the small scale motions is
responsible for producing large scale magnetic fields.
This process is explained in many text books, e.g.\ Moffatt (1978),
Parker (1979), Krause \& R\"adler (1980), Stix (2002),
or R\"udiger \& Hollerbach (2004).

The magnetic field is also responsible for linking solar variability
to natural climate variations on Earth.
Changes in the Sun's magnetic activity affect the solar irradiance by only
0.1\%, which is generally regarded as being too small to affect the climate.
However, the UV radiation is more strongly modulated and may affect the
climate.
According to an alternative proposal, the Sun's magnetic field
shields the galactic cosmic radiation, which may affect the
production of nucleation sites for cloud formation that in turn
affects the climate.
Thus, an increase in the solar field strength increases the shielding,
decreases the cosmic ray flux on Earth, decreases the cloud cover, and
hence increases the temperature.
This chain of events is rather simplified, and there can be drastically
different effects from high or low clouds, for example.
For a recent review of this rapidly developing field see
Marsh \& Svensmark (2000).

We begin by discussing the theoretical foundations governing the
properties of turbulent convection zones, and discuss then helioseismology
as an observational tool to determine, for example, the location of the
bottom of the convection zone as well as the internal angular velocity.
We turn then attention to the properties of the Sun's magnetic field and
discuss dynamo theory as its theoretical basis.
Magnetic field generation is caused both by the turbulent convection and
by the large scale differential rotation, which itself is a consequence
of turbulent convection, as will be discussed in the last section of
this chapter.
Only a bare minimum of references can be given here, and we have
to restrict ourselves mostly to reviews which give an exhaustive overview
of the original literature.
Original papers are here quoted mainly in connection with figures used in the
present text.

\section{Radial structure}

In order to determine the depth of the convection zone in the Sun
and the approximate convective velocities it is necessary to solve the
equations governing the radial structure of a star.
For this purpose the Sun can be regarded as spherically symmetric.
The equations governing the radial structure of the Sun (or a star)
are quite plausible and easily derived.
They can be written as a set of four ordinary differential equation,
namely the
\begin{itemize}
\item equation for the Sun's gravitational field (Poisson equation),
\item hydrostatic equilibrium (momentum equation),
\item thermal equilibrium (energy equation),
\item radiative equilibrium (radiation transport equation, convection).
\end{itemize}
These are given in all standard text books on stellar structure
(e.g.\ Kippenhahn \& Weigert 1990).
In the following we discuss only a subset of these equations in order
to describe some essential properties of the solar convection zone.

\subsection{Global aspects}

The rate of energy production of the Sun, i.e.\ its
luminosity, is $L_\odot=4\times10^{26}\W$
or $4\times10^{33}\erg\s^{-1}$.
The total intercepted by the Earth is only a small fraction,
\begin{equation}
  {\pi R_E^2\over 4\pi R_{E\odot}^2}=4\times10^{-10},
\end{equation}
where $R_E$ is the radius of the Earth ($6400\km$) and
$R_{E\odot}$ is the distance between the Earth and the Sun
($=1\AU=1.5\times10^{11}\m$).
Thus, the total power reaching the projected surface of the
Earth is $4\times10^{-10}\times4\times10^{26}\W=1.6\times10^{17}\W$.
This is still a lot compared with the total global energy consumption,
which was $1.4\times10^{13}\W$ in the year 2001.

The total thermal energy content of the Sun can be approximated
by half its potential energy (Virial theorem), i.e.\
\begin{equation}
E_{\rm th}\approx{GM_\odot^2\over2R_\odot}
=2\times10^{41}\J,
\end{equation}
where $G\approx7\times10^{-11}\m^3\kg^{-1}\s^{-2}$ is Newton's constant,
$M_\odot\approx2\times10^{30}\kg$ is the mass of the Sun, and
$R_\odot\approx7\times10^8\m$ is its radius.
The time it would take to use up all this energy to sustain the
observed luminosity is the Kelvin-Helmholtz time,
\begin{equation}
\tau_{\rm KH}=E_{\rm th}/L_\odot\approx10^7\yr,
\end{equation}
which is long compared with time scales we could observe directly,
but short compared with the age of the Sun and the solar system
($5\times10^9\yr$).
Therefore, gravitational energy (which is extremely efficient
in powering quasars!) cannot be the mechanism powering the Sun.
This motivated the search for an alternative explanation, which
led eventually to the discovery of the nuclear energy source of stars.

The similarity between gravitational and thermal energies
can be used to estimate the central temperature
of a star by equating $GM/R={\cal R}T_c/\mu$. For the Sun this gives
\begin{equation}
T_c\sim{\mu\over{\cal R}}\,{GM\over R}
=1.5\times10^7\K 
\label{Tc}
\end{equation}
for its central temperature.
Here, ${\cal R}\approx8300\m^2\s^{-2}\K^{-1}$ is the universal gas
constant and $\mu\approx0.6$ is the non-dimensional mean molecular weight
for a typical mixture of hydrogen and helium.
The estimate \eq{Tc} happens to be surprisingly accurate.
This relation also tells us that the central temperature of the Sun
is only determined by its mass and radius, and not, as one might have
expected, by the luminosity or the effectiveness of the nuclear reactions
taking place in center of the Sun.

\subsection{Thermal and hydrostatic equilibrium}

The condition of hydrostatic equilibrium can be written in the form
\EQ
0=-{1\over\rho}\nab p+\grav,
\label{HydrostaticSupport}
\EN
where $\rho$ is the density, $p$ is the pressure, and $\grav$ is the
gravitational acceleration.
In the spherically symmetric case we have $\grav=-(GM_r/r^2,0,0)$ in spherical
polar coordinates, where $M_r$ is the mass inside a sphere of radius $r$.
\EEq{HydrostaticSupport} is readily solved in the special case
where the radial dependence of the density is polytropic, i.e.\
$\rho(r)\sim T(r)^m$, where $m$ is the polytropic index.
This yields
\EQ
T(r)=T_c-{1\over1+m}{\mu\over{\cal R}}\int_0^r{GM_r\over r^2}\,\dd r.
\label{Teqn}
\EN
So, in the outer parts of the Sun, where $M_r\approx\const$,
\Eq{Teqn} can be integrated, which shows that the temperature
has a term that is proportional to $1/r$.

Significant amounts of energy can only be produced in the inner
parts of the Sun where the temperatures are high enough for
nuclear reactions to take place.
The central temperature is characterized by the condition of thermal
energy equilibrium, which quantifies the rate of change of the local
luminosity, $L_r$, with radius.
Outside the core, nuclear reactions no longer take place, so $L_r$ can be
considered constant.
The radiative flux is given by $F=L_r/(4\pi r^2)$, which thus
decreases like $1/r^2$ in the outer parts.

In the bulk of the Sun, energy is transported by photon
diffusion: the optical mean-free path is short compared with other
relevant length scales (e.g.\ pressure scale height), so we are in the
optically thick limit and can use the diffusion approximation for photons.
The radiative flux, $F$, is therefore in the negative direction of
and proportional to the gradient of the radiative energy density, $aT^4$,
where $a=7.57\times10^{-15}\erg\cm^{-3}\K^{-4}$ is the
radiation-density constant.
The connection between fluxes and concentration gradients is generally
referred to as Fickian diffusion.
As in kinetic gas theory, the diffusion coefficient is 1/3 times the
typical particle velocity (=speed of light $c$) and the mean free path
$\ell$ of the photons, so
\begin{equation}
F=-\onethird c\ell\;{\dd\over\dd r}(aT^4)
=-\fourthird ac\ell T^3\,{\dd T\over\dd r}
\approx-K\,{\dd T\over\dd r},
\label{RadDiffApprox}
\end{equation}
which is basically the condition of radiative equilibrium.
Here we have introduced the radiative conductivity $K$.
The photon mean free path is usually expressed in terms of the opacity
$\kappa$, which is the effective cross-section per unit mass, so
$\ell=(\rho\kappa)^{-1}$.
Expressing $a$ in terms of the Stefan-Boltzmann constant,
$\sigma_{\rm SB}=ac/4$, we have
\EQ
K={16\sigma_{\rm SB}T^3\over3\kappa\rho}.
\label{Keqn}
\EN
An approximation for the opacity $\kappa$ that is commonly used for
analytic considerations is Kramer's formula
\begin{equation}
  \kappa = \kappa_0\rho T^{-7/2}\quad\mbox{(Kramer's opacity)},
\end{equation}
where $\kappa_0=6.6\times10^{18}\m^5\K^{7/2}\kg^{-2}$ for so-called
free-free transitions where two charged particles form a system which
can absorb and emit radiation.
This value may well be up to 30 times larger if the gas is rich in heavier
elements, so it is a good electron supplier and bound-free processes
(ionization of neutral hydrogen by a photon) become important as well.
In practice, a good value is $\kappa_0\approx10^{20}\m^5\K^{7/2}\kg^{-2}$
(corresponding to $10^{24}\cm^5\K^{7/2}\g^{-2}$).
With Kramer's formula, the conductivity is
\begin{equation}
  K = {16\sigmaSB T^{13/2}\over3\kappa_0\rho^2}.
\end{equation}
For a polytropic stratification, i.e.\ when the density is given by a
power law of the temperature, $\rho\sim T^m$, we have
\begin{equation}
  K\sim T^{13/2-2m},
\end{equation}
which is constant for an effective polytropic index $m=13/4=3.25$.
This gives indeed a reasonable
representation of the stratification of stars in convectively
stable regions throughout the inner parts of the Sun.
At the bottom of the solar convection zone the density is about
$200\kg\m^{-3}$ and the temperature is about $2\times10^6\K$. This gives
$K=(3...100)\times10^9\kg\m\s^{-3}\K^{-1}$. In order to carry the solar
flux the average temperature gradient has to be around $0.01\K/\m$.

\subsection{Transition to adiabatic stratification}

In reality $K$ does change slowly with height.
Therefore the polytropic index effectively changes with height.
If $m<13/4$, then $K$ decreases with decreasing $T$.
However, in order to transport the required energy flux,
the temperature gradient has to increase, so the polytropic
index decreases further, until it reaches a critical value
where the specific entropy gradient reverses sign.
This leads to the onset of Rayleigh-Benard convection.

Specific entropy is an important quantity, because it does not change in
the absence of local heating or cooling processes.
For a perfect gas, and ignoring partial ionization effects, the specific
entropy can be defined, up to a constant $s_0$, as
\EQ
s=c_v\ln p-c_p\ln\rho+s_0,
\label{entropy}
\EN
where $c_p$ and $c_v$ are the specific heats at constant pressure
and constant volume, respectively.
Their ratio is $\gamma=c_p/c_v$, which is $5/3$ for a monatomic
gas, and their difference is $c_p-c_v={\cal R}/\mu$.

\begin{figure}[t!]\begin{center}
\includegraphics[width=.6\textwidth]{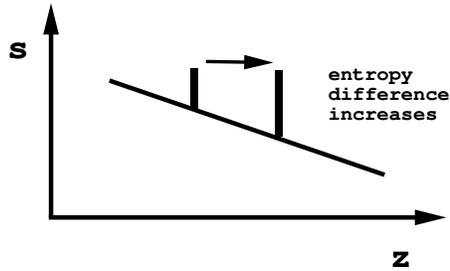}
\end{center}\caption[]{
Specific entropy profile for an unstable atmosphere.
The difference in specific entropy between the blob and the
surroundings increases as the blob ascends.
Gravity points in the negative $z$ direction, so $\grav\cdot\nab s>0$
in this case.
}\label{Fentropy}\end{figure}

If the specific entropy of the environment decreases in the upward direction,
an upward moving blob of gas will develop excess entropy; see \Fig{Fentropy}.
Assuming pressure equilibrium across the blob,
\Eq{entropy} shows that a positive
entropy excess $\delta s$ corresponds to a density deficit,
$-c_p\delta\ln\rho$.
Thus, the blob will be lighter than its surroundings and will therefore
be buoyant, which drives the convection.
Likewise, a downward moving blob will become heavier and fall even faster.

Using an equation of state for a perfect gas, i.e.\
$p=({\cal R}/\mu)\rho T$ we have $\dd\ln p=\dd\ln\rho+\dd\ln T$,
and therefore \Eq{entropy} gives
\EQ
{\mu\over{\cal R}}{\dd s\over\dd\ln T}={1\over\gamma-1}-m.
\EN
This shows that, once $m$ drops below $1/(\gamma-1)=1.5$, specific
entropy decreases in the upward direction, i.e.\ in the direction of
decreasing temperature.
As a result, convection sets in which rapidly mixes the gas and causes
the specific entropy to be nearly constant, keeping the effective value
of $m$ always close to the critical value of 1.5.

\begin{figure}[t!]\begin{center}
\includegraphics[width=\textwidth]{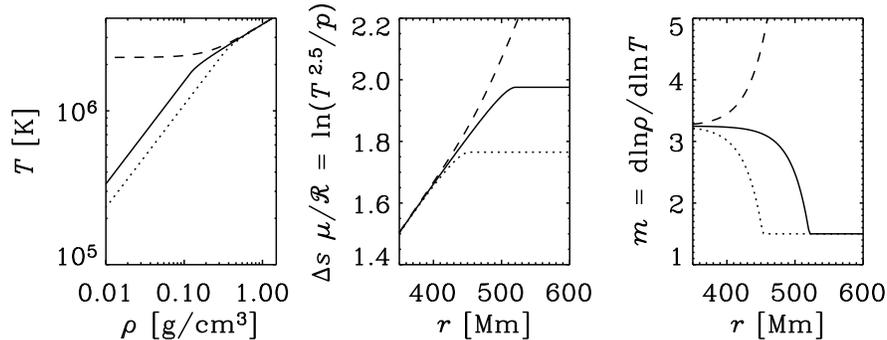}
\end{center}\caption[]{
Solutions of \Eqs{dlnpdrdlnTdr}{nabla}, starting the integration
at $r=350\Mm$ with $T=4\times10^6\K$ and three different values of the
density: $\rho=1.340$ (dashed line), $\rho=1.345$ (solid line),
and $\rho=1.350$ (dotted line).
The left hand panel shows temperature versus density.
In this panel the integration goes from right to left (i.e.\ in the
upward direction toward lower density).
The middle panel shows the radial specific entropy profile; note that
for the two cases with $\rho\leq1.345$ (solid and dotted lines) a
convection zone develops at $r\approx500$ and $420\Mm$, respectively.
These two cases correspond to cases where $m<13/4$ at the lower
boundary, as seen from the right hand panel.
Note the positive entropy gradient indicating stability.
In the last two panels the integration goes from left to right.
}\label{pcomp}\end{figure}

In order to calculate the actual stratification, we need to solve
\Eqs{HydrostaticSupport}{RadDiffApprox} together with the equations
describing the increase of $M_r$ and $L_r$ with radius.
Assuming that $M_r$ and $L_r$ are constant (valid far enough
away from the core), we are left with two
equations, which we express in terms of $\ln p$ and $\ln T$, so
\EQ
{\dd\ln p\over\dd r}=-{\mu\over{\cal R}T}\,{GM_r\over r^2},
\label{dlnpdr}
\EN
\EQ
{\dd\ln T\over\dd r}=-{1\over KT}\,{L_r\over4\pi r^2}.
\label{dlnTdr}
\EN
It is convenient to integrate these equations in the form
\EQ
{\dd\ln p\over\dd r}=-{1\over H_p}\quad\mbox{and}\quad
{\dd\ln T\over\dd r}=-{\nabla\over H_p},
\label{dlnpdrdlnTdr}
\EN
where the symbol $\nabla$ is commonly used in astrophysics for
the local value of $\dd\ln T/\dd\ln p$, and $H_p={\cal R}T/(\mu g)$
is the local pressure scale height.
In the convectively stable regions, i.e.\ where $m>3/2$
(corresponding to $\nabla<\nabla_{\rm ad}=2/5$, and neglecting partial
ionization effects), we have
$\nabla=\nabla_{\rm rad}$, where $\nabla_{\rm rad}$ can be found by
dividing \eq{dlnTdr} by \eq{dlnpdr}, so
\EQ
\nabla_{\rm rad}={1\over K}\,{{\cal R}\over\mu}\,{L_r\over4\pi GM_r}.
\label{nabla}
\EN
Inside convection zones, on the other hand, $\nabla$ is replaced by
$\nabla_{\rm ad}$, so in general we can write
$\nabla=\min(\nabla_{\rm rad},\nabla_{\rm ad})$.
In \Fig{pcomp} we show solutions obtained by integrating from $r=500\Mm$
(1\,Mm=1000\,km)
upward using $T=4\times10^6\K$ as starting value with $\rho$ chosen
such that the resulting value of $m$ is either just below or just above
$13/4=3.25$.

The considerations above have demonstrated that $m$ must indeed be
quite close to $13/4=3.25$ in the radiative interior, but that its value
decreases over a depth of about $50\Mm$ to the adiabatic value of 1.5
just below the bottom of the convection zone.
The precise location of the bottom of the convection zone depends on
the value of specific entropy in the bulk of the convection zone; see
the middle panel of \Fig{pcomp}.
This value depends on the detailed surface physics and in particular the
value of the opacity at the top of the convection zone.
Here the Kramers opacity is no longer appropriate and the opacity from
producing a negative hydrogen ion by polarizing a neutral hydrogen atom
through a nearby charge becomes extremely important.

\subsection{Mixing length theory and convection simulations}

The approximation of setting $\nabla=2/5$ in the unstable region
becomes poor near the surface layers where density is small
and energy transport by turbulent elements less efficient.
In fact, if the specific entropy were completely constant throughout
the convection zone, there would be no net exchange of entropy
by the turbulent elements.
The definition for the convective flux is
\begin{equation}
\FF_{\rm conv}=\overline{(\rho\uu)'c_p T'},
\label{conv}
\end{equation}
where overbars denote horizontal averages and primes denote
fluctuations about these averages.
A mean field calculation shows that $\FF_{\rm conv}$ is proportional to
the negative entropy gradient (see the monograph by R\"udiger 1989),
\begin{equation}
\FF_{\rm conv}=-\chi_{\rm t}\,\meanrho\,\meanT\,\nab\,\means
\quad\quad\mbox{(if $\grav\cdot\nab\,\means>0$)},
\label{grads}
\end{equation}
where $\chi_{\rm t}$ is a turbulent diffusion coefficient.
In the following we omit the overbars for simplicity.
Note that, by comparison with \Eq{RadDiffApprox}, in a turbulent
environment Fickian diffusion down the {\it temperature} gradient gets
effectively replaced by a similar diffusion down the {\it entropy} gradient.
As with all other types of diffusion coefficients, the diffusion
coefficient is proportional to the speed of the fluid parcels
accomplishing the diffusion, and the length over which such
parcels stay coherent (i.e.\ the mean free path which is commonly
also denoted as the mixing length).
Thus, we have
\begin{equation}
\chi_{\rm t}=\onethird u_{\rm rms}\ell.
\label{chi}
\end{equation}
The subscript ${\rm t}$ indicates that this coefficient applies to
turbulent transport of averaged fields.
Given that the total flux is known, and also the fractional contribution
from the radiative flux, we know also the convective flux.
Thus, \Eq{grads} can be used to determine the radial entropy gradient,
provided we know $\chi_{\rm t}$, and hence $u_{\rm rms}$ and $\ell$.

\begin{figure}[t!]\includegraphics[width=\textwidth]{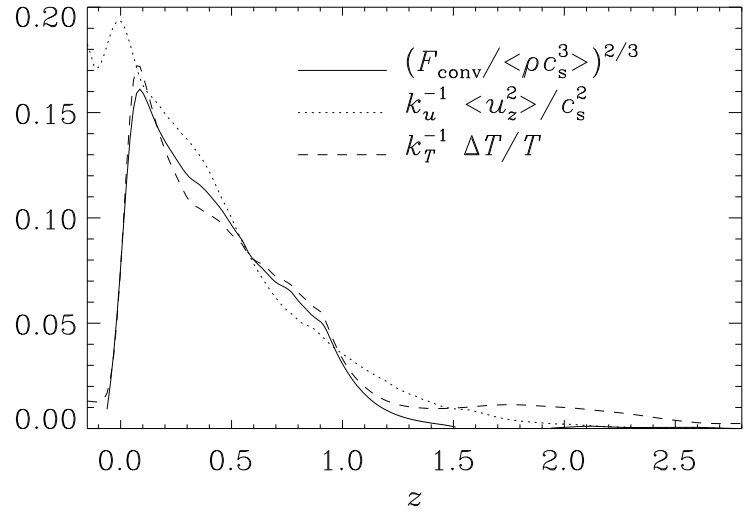}\caption[]{
Vertical profiles of the normalized mean squared vertical velocity
fluctuations and temperature fluctuations, compared with the normalized
convective flux raised to the power $2/3$. Note the good agreement
between the three curves within the convection zone proper.
In this plot, $z$ denotes depth.
The positions $z=0$ and 1 corresponds to the top and bottom of the
convection zone, respectively.
There is a lower overshoot layer for $z>1$ and an upper overshoot
layer for $z<0$.
[Adapted from Brandenburg et al.\ (2005).]
}\label{Fprfluctn}\end{figure}

A natural length scale in the problem is the
scale height, so we assume that the mixing length is some fraction
$\alpha_{\rm mix}$ of the local vertical pressure scale height, i.e.\
\begin{equation}
\ell=\alpha_{\rm mix} H_p.
\label{mix}
\end{equation}
The scaling of the rms velocity is constrained by \Eq{conv}.
Assuming that temperature and velocities are well correlated
(warm always up, cool always down), we can also write
\begin{equation}
F_{\rm conv}\approx\rho u_{\rm rms} c_p\delta T,
\end{equation}
where $\delta T=(\overline{T'^2})^{1/2}$ is the rms temperature
fluctuation.
The relative proportion, with which convection produces velocity and
temperature fluctuations, can be estimated by balancing
the buoyancy force of a blob against its drag force, so 
$F_{\rm buoy}=F_{\rm D}^{\rm(turb)}$ and therefore
$\delta\rho\, gV = C_{\rm D}\rho u_{\rm rms}^2S$,
where $C_{\rm D}$ is the drag coefficient,
$V$ is the volume of the blob, and $S$ its cross-sectional area.
We parameterize the ratio $V/(C_{\rm D}S)=\alpha_{\rm vol}H_p$,
where $\alpha_{\rm vol}$ is a nondimensional factor of order unity
characterizing the blob's volume to surface ratio.
Assuming pressure equilibrium we have furthermore
$|\delta\rho/\rho|=|\delta T/T|$, and
\begin{equation}
u_{\rm rms}^2=\alpha_{\rm vol}\,g H_p\,{\delta T\over T}.
\label{buoyancy}
\end{equation}
Thus, $u_{\rm rms}$ is proportional to $(\delta T/T)^{1/2}$, so
$F_{\rm conv}$ is proportional to $(\delta T/T)^{3/2}$, and therefore
\begin{equation}
\delta T/T\sim F_{\rm conv}^{2/3}\quad\mbox{and}\quad
u_{\rm rms}/\cs\sim F_{\rm conv}^{1/3}.
\end{equation}
These scaling relations hold also locally at each depth; see
\Fig{Fprfluctn}, where we show that in simulations of Rayleigh-Benard
convection; the vertical profiles of the normalized mean squared
vertical velocity, $\bra{u_z^2}/c_{\rm s}^2$, and of the relative
temperature variance, $\delta T/T$, are indeed locally proportional
to $[F_{\rm conv}/(\rho c_{\rm s}^3)]^{2/3}$.
In \Fig{Fprfluctn} the nondimensional coefficients are
$k_T\approx1.1$ and $k_u\approx0.4$, which implies
\EQ
F_{\rm conv}\approx k_u^{-3/2}\rho u_{\rm rms}^3,
\label{Fconvurms}
\EN
where $k^{-3/2}\approx4$ and $\bra{u_z^2}=u_{\rm rms}^2$ has been used.
Using $F_{\rm conv}=7\times10^7\W\m^{-2}$ and $\rho=10\kg\m^{-3}$ at a
depth of about $40\Mm$ this equation implies $u_{\rm rms}=120\m\s^{-1}$.

Using \Eq{grads} and the fact that
$\chi_{\rm t}\propto u_{\rm rms}\propto F_{\rm conv}^{1/3}$ we have
$F_{\rm conv}\propto F_{\rm conv}^{1/3}|ds/dz|$, or\footnote{
A more rigorous calculation using the equations above shows that
$${ds/c_p\over dz}=-{k\over H_p}
\left({F_{\rm tot}\over\rho \cs^3}\right)^{2/3},
\quad\mbox{where}\quad
k=3\,{\gamma-1\over\alpha_{\rm mix}}
\left[\alpha_{\rm vol} \left(1-{1\over\gamma}\right)\right]^{-1/3},$$
and $k\approx1$ for $\alpha_{\rm mix}=\alpha_{\rm vol}=2$.
}
\begin{equation}
|ds/dz|\propto F_{\rm conv}^{2/3}.
\end{equation}
In calculating the specific entropy gradient, we can, as a first approximation,
assume that $F_{\rm conv}$ is approximately the total flux.
However, it would not be difficult to calculate the entropy gradient
self-consistently by solving a cubic equation.
We also note that the entropy gradient is related to $\nabla$ by
\EQ
{\dd s/c_p\over\dd\ln p}=\nabla-\nabla_{\rm ad},\quad\mbox{where}
\quad\nabla_{\rm ad}=1-{1\over\gamma}.
\EN
A solution of the full system of equations, which include more realistic
physics than what has been described here, has been given by
Spruit (1974); see \Tab{SolarModel}.
The rms velocities are about half as big as expected from \Eq{Fconvurms}.

\begin{table}[b!]\caption{The solar mixing length model of Spruit (1974).}
\vspace{12pt}\centerline{\begin{tabular}{ccccccccc}
$\quad z\,\mbox{[Mm]}$ & $\quad T\,\mbox{[K]}$ &
$\quad\rho\,\mbox{[g\,cm$^{-3}$]}$ & $\quad H_p\,\mbox{[Mm]}$ &
$\quad u_{\rm rms}\,\mbox{[m/s]}$ & $\quad\tau\,\mbox{[d]}$ &
$\quad\nu_{\rm t}\,\mbox{[cm$^2$/s]}$ & $\quad\Omega_0\tau$ \\
\hline
 24 & $1.8\times10^5$ & 0.004 &  8 & 70 & 1.3 & $1.5\times10^{12}$ & 0.6 \\
 39 & $3.0\times10^5$ & 0.010 & 13 & 56 & 2.8 & $2.0\times10^{12}$  & 1.3 \\
155 & $1.6\times10^6$ & 0.12  & 48 & 25 &  22 & $3.2\times10^{12}$ & 10 \\
198 & $2.2\times10^6$ & 0.20  & 56 &  4 & 157 & $0.6\times10^{12}$ & 70 \\
\label{SolarModel}\end{tabular}}\end{table}

Near and beyond the upper and lower
boundaries of the convection zones the approximation \eq{buoyancy}
becomes bad, because it ignores the fact that convective elements
have inertia and can therefore overshoot a significant distance into
the stably stratified regions. In those layers where the entropy gradient
has reversed, a downward moving fluid parcel becomes hotter than its
surroundings.
Thus, in these layers the convection
carries convective flux downward, so its sign is reversed.
Simulations have clearly demonstrated that,
owing to strong stratification, convection will be highly
inhomogeneous, with narrow downdrafts and broad upwellings.
This leads to a characteristic (but irregular) pattern of
convection; see, e.g., the text book by Stix (2002).

The precise location of the bottom of the convection zone in now
fairly well determined from detailed models of stellar structure,
where the full evolution from a zero-age main sequence star to a
chemically evolved star where some of the hydrogen has been burnt
into helium and other elements, has been taken into account.
An even more accurate and quite independent determination of
the bottom of the convection zone and the overall stratification
is possible through helioseismology.
This will be discussed in the next section.

\section{Helioseismology}
\label{Shelioseis}

The Sun exhibits so-called five-minute oscillations that are best seen
in spectral line shifts.
These oscillations were first thought to be the oscillatory response
of the atmosphere to convection granules pushing upwards into stably
stratified layers.
This idea turned out to be wrong, because the oscillations are
actually {\it global} oscillations penetrating deep layers of the Sun.
In fact, they are just sound waves that are {\it trapped}
in a cavity formed by reflection at the top
and refraction in deeper layers.
At the top, sound waves cannot penetrate if their wave length
exceeds the scale on which density changes.
The refraction in deeper layers is caused by the higher wave speed
of the wave front in those parts that are deeper in the Sun.
This makes the wave front bend back up again.

\begin{figure}[t!]\begin{center}
\includegraphics[width=\textwidth]{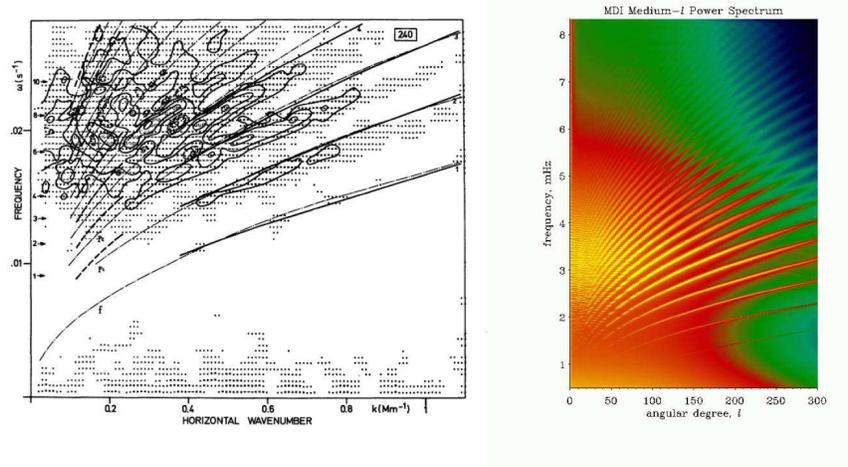}
\end{center}\caption[]{
Comparison between the $k_{\rm h}$--$\omega$ (or $l$--$\nu$) diagrams
obtained by Deubner in 1975 (left), and by the SOHO/MDI team in 2000 (right).
The figure by Deubner, where he compares observations with the predictions
by Ulrich (1970, solid lines), proved that the 5-min oscillations
were global modes.
Courtesy F.-L.\ Deubner (left) and P.\ Scherrer (right).
}\label{deubner}\end{figure}

The decisive observation came when a wavenumber--frequency
(or $k$--$\omega$) diagram was produced that showed that these modes
have long term and large scale spatio-temporal coherence with wavenumbers
corresponding to 20--60 Mm; see \Fig{deubner}.

By now the determination of $k$--$\omega$ diagrams
has grown to a mature and standard tool in solar physics.

\subsection{Qualitative description}

Since the beginning of the eighties, standing acoustic waves in the Sun
have been studied in great detail.
It has become possible to measure directly (i.e.\ without the
use of a solar model)
\begin{itemize}
\item[(i)] the radial dependence of the sound speed, $\cs(r)$, which is
proportional to the temperature.
Note that $\cs^2=\gamma p/\rho=\gamma{\cal R}T/\mu$, but the mean
molecular weight increases near the core due to the nuclear reaction
products.
\item[(ii)] the radial and latitudinal dependence of the internal angular
velocity, $\Omega=\Omega(r,\theta)$, throughout the Sun.
\end{itemize}
This technique is called {\it helioseismology}, because it is
mathematically similar to the techniques used in seismology of the
Earth's interior.
Qualitatively, the radial dependence of the sound speed can be 
measured, because standing sound waves of different horizontal wave 
number penetrate to different depths. Therefore, the frequencies of those
different waves depend on how exactly the sound speed changes with depth.
Since the Sun rotates, the waves that travel in the direction of 
rotation (i.e.\ toward us) are blue-shifted, and those that travel
against the direction of rotation (i.e.\ away from us) are red-shifted.
Therefore, the frequencies are split,
depending on the amount of rotation in different layers.
There are many reviews on the subject (e.g., Demarque \& Guenther 1999).
Here we follow the text book by Stix (2002).

Acoustic waves are possible, because they are constantly being
excited by the ``noise'' generated in the convection zone via
stochastic excitation. The random
fluctuations in the convection are turbulent and contain power at all
frequencies.
Now the Sun is a harmonic oscillator for sound waves and the
different sound modes can be excited stochastically.

Helioseismology has now grown to be immensely sophisticated and more accurate
data have emerged from observations with the Michelson Doppler Imager
aboard the SOHO spacecraft, located at the inner Lagrange
point between Sun and Earth, and also the GONG project
(GONG = Global Oscillation Network Group).
The latter involves six stations around the globe to eliminate
nightly gaps in the data.

\subsection{Inverting the frequency spectrum}

As with a violin string, the acoustic frequency of the wave increases
as the wavelength decreases.
More precisely, the frequency is given by $\nu=\cs/\lambda$,
where $\lambda$ is the wavelength and $\cs$ is the sound speed.
We will also use the circular frequency $\omega=2\pi\nu$ with
$\omega = \cs k$, where $k=2\pi/\lambda$ is the wavenumber.
If sound waves travel an oblique path then we can express the wavenumber
in terms of its horizontal and vertical wavenumbers, $k_{\rm h}$ and
$k_{\rm v}$, respectively. We do this because only the horizontal
wavenumber can be observed. This corresponds to the horizontal pattern in
\Fig{deubner}. Thus, we have
\begin{equation}
  k^2 = k_{\rm h}^2+k_{\rm v}^2.
\end{equation}
The number of radial nodes of the wave is given by the number of waves
that fit into that part of the Sun where the corresponding wave can travel.
This part of the Sun is referred to as the cavity.
The larger the cavity, the more nodes there are for a given wavelength.
The number of modes $n$ is given by
\begin{equation}
  n = 2\Delta r/\lambda
 = 2\Delta r {k_{\rm v}\over2\pi}
 = \Delta r\,k_{\rm v}/\pi,
\end{equation}
where $\Delta r$ is the depth of the cavity.
If the sound speed and hence $k_{\rm v}$ depend on radius, this formula
must be generalized to
\begin{equation}
  n = \frac{1}{\pi}\int\limits_{r_{\rm min}}^{R_\odot}k_{\rm v}\,dr,
\end{equation}
supposing the cavity to be the spherical shell
$r_{\rm min} < r < R_\odot$.

\begin{figure}[t!]\begin{center}
\includegraphics[width=.8\textwidth]{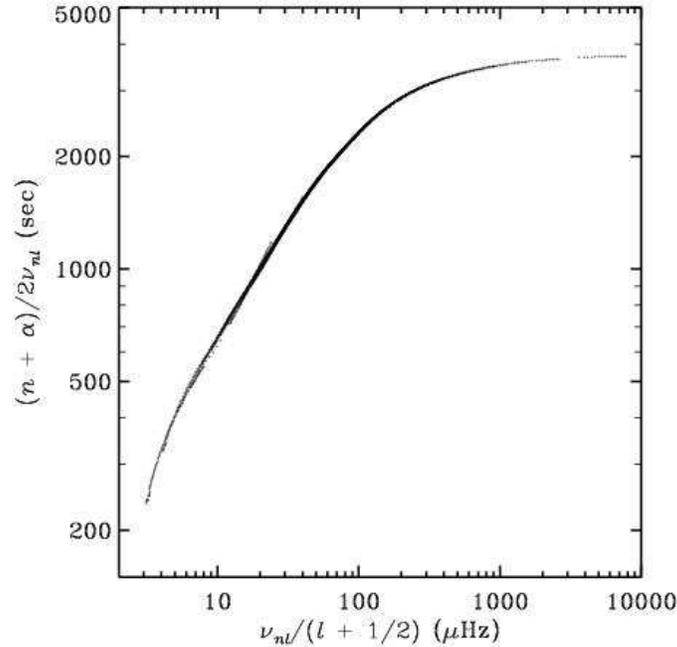}
\end{center}\caption[]{
The Duvall law.
The vertical axis (ordinate) corresponds to $F$ in \Eq{Helioseis-Ieq}
and the horizontal axis (abscissa) is basically $u^{-1}$.
He found this law well before its significance was understood in terms
of one of the functions in Abel's integral transformation.
[Courtesy J.\ Christensen-Dalsgaard et al.\ (1985).]
}\label{duvall-law}\end{figure}

The horizontal pattern of the proper oscillation is described by spherical
harmonics with indices $l$ and $m$, hence the horizontal wavenumber is
\begin{equation}
  k_{\rm h}^2 = \frac{\ell(\ell+1)}{r^2},
\end{equation}
and we can write
\begin{equation}
  k_{\rm v} = \sqrt{\frac{\omega_{nl}^2}{\cs^2}
  - \frac{\ell(\ell{+}1)}{r^2}}
  = \frac{\omega_{nl}}{r}\sqrt{\frac{r^2}{\cs^2}
  - \frac{\ell(\ell{+}1)}{\omega_{nl}^2}}.
\end{equation}
where the subscripts of $\omega_{nl}$ denote the radial order $n$ and the
spherical harmonic degree $l$ of the modes.
Therefore, the number $n$ of radial nodes is given by 
\begin{equation}
  \frac{\pi(n+\alpha)}{\omega_{nl}}
  =\int\limits_{r_{\rm min}}^{R_\odot}
  \sqrt{\frac{r^2}{\cs^2}-\frac{\ell(\ell{+}1)}{\omega_{nl}^2}}\frac{dr}{r},
\label{RadialNodes}
\end{equation}
where an empirically (or otherwise) determined
phase shift $\alpha\approx 1.5$ accounts for the fact that the
standing waves are confined by barriers that are ``soft'' and extended,
rather than rigid and fixed.

The location of the inner turning radius is given by the point where
the wavevector has become completely horizontal.
Using
\EQ
\omega_{nl}^2/\cs^2=k^2=k_{\rm h}^2+k_{\rm v}^2,
\EN
together with $k_{\rm v}=0$ at $r=r_{\min}$ and $k_{\rm h}^2=\ell(\ell+1)/r^2$,
we have $(r_{\rm min}/\cs)^2=\ell(\ell+1)/\omega_{nl}^2$.
This implies that
\begin{equation}
r_{\rm min}={\cs\over\omega_{nl}}\sqrt{\ell(\ell+1)},
\end{equation}
so only modes with low $\ell$ values have turning points close to the
center and can be used to examine the Sun's core.
We now introduce new variables
\begin{equation}
  \xi = \frac{r^2}{\cs^2}, \qquad\quad
  u = \frac{\ell(\ell{+}1)}{\omega_{nl}^2},
\end{equation}
so the inner turning point of the modes corresponds to $\xi=u$.
Furthermore, we denote the left hand side of \Eq{RadialNodes}
by $F(u)$, so we can write
\begin{equation} \label{Helioseis-Ieq}
  F(u) = \int\limits_{u}^{\xi_\odot}\sqrt{\xi-u}\,\frac{d\ln r}{d\xi} \,d\xi,
\end{equation}
where the location of the inner refraction point corresponds to $u=\xi$.
The function $F(u)$ was obtained from observations by
Duvall (1982) on the grounds that
this combination of data makes the different branches collapse
onto one (see \Fig{duvall-law}).
He discovered this well before its significance was understood by
Gough (1985) several years later.

\begin{figure}[t!]\begin{center}
\includegraphics[width=.5\textwidth]{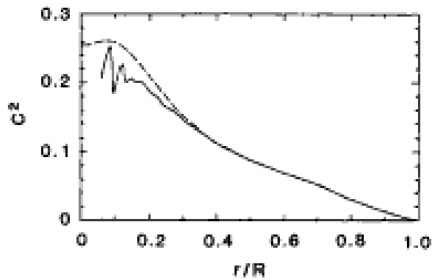}
\end{center}\caption[]{
Radial dependence of the sound speed on radius in the Sun. Note the
change in slope near a radius of $0.7$ solar radii. The oscillations near the
center are not physical. The theoretical model (dotted line) is in
fair agreement with the direct measurements. The sound speed has its
maximum not in the center, because the mean molecular weight $\mu$
increases towards the center, which causes $\cs$ to decrease. 
(We recall that $\cs^2(r)=\gamma{\cal R}T/\mu$.)
[Adapted from Stix (2002).]
}\label{Fsound}\end{figure}

Since we know $F(u)$ from observations and are
interested in the connection between $r$ and $\xi$ (i.\,e. $r$ and $\cs$),
we interpret \eq{Helioseis-Ieq} as an integral equation for the unknown
function $r(\xi)$.
Most integral equations cannot be solved in closed form, but this one can.
Gough (1985) realized that it can be cast in the form of
Abel's integral equation.
The pair of complementary equations (primes denote derivatives) is
\begin{equation} \label{Helioseis-Ieq2}
F(u) = \int\limits_{u}^{\xi_\odot}\sqrt{\xi-u}\;G'(\xi)\,d\xi,
\label{Abel1}
\end{equation}
\begin{equation} \label{Helioseis-Ieq3}
G(\xi) = {2\over\pi}\int\limits_{\xi}^{\xi_\odot}{1\over\sqrt{\xi-u}}\,
F'(u)\,du.
\label{Abel2}
\end{equation}
Inserting the definitions for $\xi$ and $u$ into \Eq{Abel2}, we obtain
\begin{equation}
\int\limits_{\xi}^{\xi_\odot} \frac{F'(u)}{\sqrt{u-\xi}} \,du
= -\frac{\pi}{2}\int\limits_{\xi}^{\xi_\odot} \frac{d\ln r}{d\xi'}\,d\xi'
= -\frac{\pi}{2} \left.\rule{0pt}{2.5ex} \ln r \right|_{\xi'=\xi}^{\xi_\odot}
= \frac{\pi}{2} \ln \frac{r(\xi)}{R_\odot}.
\end{equation}
This equation can be solved for $r=r(\xi)$:
\begin{equation}
r(\xi) = R_\odot \exp\Biggl( \frac{2}{\pi}\int\limits_{\xi}^{\xi_\odot}
                            \frac{F'(u)}{\sqrt{u-\xi}}\,du \Biggr).
\end{equation}
This is the final result of inverting the integral equation
(\ref{Helioseis-Ieq}). It establishes the link between
the observable function $F(u)$ and the function $r(\xi)$, from which the
radial profile of the sound velocity $\cs$ can be obtained.
\FFig{Fsound} gives the result of an inversion procedure that computes the
radial dependence of the sound speed on depth, using the detailed
frequency spectrum as input.

It should be noted, however, that this approach is usually not practical
when input data are noisy.
Instead, a minimization procedure is often used where the
resulting function is by construction smooth.
This procedure falls under the general name of {\it inverse theory} and
is frequently used in various branches of astrophysics.

Historically, the model independent determination of the sound speed
and thereby the temperature in the center of the Sun has been important
in connection with understanding the origin of the solar neutrino
problem.
In fact, the solar neutrino flux was measured to be only one third of
that originally expected.
A lower core temperature could have resolved this mismatch, but this
possibility was then ruled out by helioseismology.
Now we know that there are neutrino oscillations leading to a continuous
interchange between the three different neutrino species, which explains
the observed neutrino flux of just one species.

\subsection{The solar abundance problem}
\label{CurrentProblems}

Opacities depend largely on the abundance of heavier elements.
The solar models calculated with the old tables agreed quite well
using the conventional abundance ratio of heavier elements to hydrogen,
$Z/X=0.023$.
However, the abundancies were based on fits of observed spectra
to synthetic line spectra calculated from model atmospheres.
These models parameterize the three-dimensional convection only rather crudely.
New synthetic line spectra calculated from three-dimensional time-dependent
hydrodynamical models of the solar atmosphere give a lower value of
the solar oxygen abundance.
With the new values ($Z/X=0.017$) it became difficult to reconcile
the previously good agreement between stellar models and helioseismology.
The solution to this problem is still unclear, but there is now
evidence that the solar neon abundance may have been underestimated.
A neon abundance enhanced by about 2.5 is sufficient to restore the
good agreement found previously.

The detailed stratification depends quite sensitively on the equation
of state, $p=p(\rho,T)$.
However, the uncertainties in the theoretically determined equation of
state are now quite small and cannot be held responsible for reconciling
the helioseismic mismatch after adopting the revised solar abundancies.

\subsection{Internal solar rotation rate}

\begin{figure}[t!]\centering
\includegraphics[width=\textwidth]{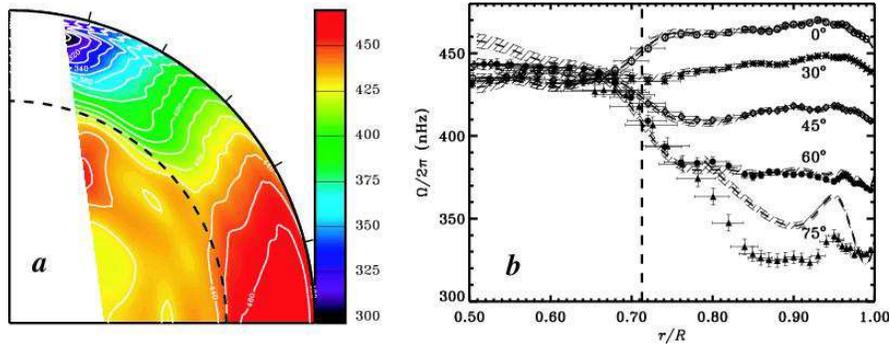} 
\caption[]{ 
Angular velocity profile in the solar interior inferred from
helioseismology (after Thompson et al.\ 2003).
In panel (a) a two-dimensional (latitude-radius) rotational inversion is
shown based on the Subtractive Optimally Localized Averaging (SOLA)
technique. In panel (b) the angular velocity is plotted as a function
of radius for several selected latitudes, based on both SOLA (symbols,
with $1\sigma$ error bars) and regularized least squares (RLS; dashed lines)
inversion techniques. Dashed lines indicate the base of the convection
zone. All inversions are based on data from the Michelson Doppler
Imager aboard the SOHO spacecraft, averaged over
144 days. Inversions become unreliable close to the rotation axis,
represented by white areas in panel (a). Note also that global modes
are only sensitive to the rotation component which is symmetric about
the equator (courtesy M.J. Thompson et al.\ 2003).
}\label{global1}
\end{figure}

Another important problem is to calculate the
internal rotation rate of the Sun (\Fig{global1}).
This has already been possible for the past 20 years,
but the accuracy has been ever improving.
We will not discuss here the mathematics in any further detail,
but refer instead to the review by Thompson et al.\ (2003).
The basic technique involves the prior calculation of kernel functions,
$K_{nlm}(r,\theta)$, that are independent of $\Omega$, such that
the rotational frequency splitting can be expressed as
\EQ
\omega_{nlm}-\omega_{nl0}=m\int_0^R\int_0^\pi
K_{nlm}(r,\theta)\Omega(r,\theta)r\dd r\dd\theta
\EN
Several robust features that have emerged from the work of several
groups include
\begin{itemize}
\item The contours of constant angular velocity do not show a tendency of
alignment with the axis of rotation, as one would have expected, and as
many theoretical models still show.
\item The angular velocity in the radiative interior is
nearly constant, so there is no rapidly rotating core,
as has sometimes been speculated.
\item There is a narrow transition layer at the bottom of the
convection zone, where the latitudinal differential rotation goes
over into rigid rotation (i.e.\ the tachocline).
Below $30^\circ$ latitude the radial angular velocity gradient is
here positive, i.e.\ $\partial\Omega/\partial r>0$, in contrast
that what is demanded by conventional dynamo theories.
\item Near the top layers (outer 5\%) the angular velocity
gradient is negative and quite sharp.
\end{itemize}

A completely model-independent knowledge about the internal rotation
rate of the Sun has proved to be invaluable for the theory of the magnetic
field in the Sun, for its rotation history, and for solar dynamo theory.
Prior to the advent of helioseismology some 25 years ago,
the idea of a rapidly rotating core was quite plausible,
because at birth the Sun is believed to have spun at least 50 times
faster than now, and because in the Sun the viscous time scale exceeds
the age of the Sun.
The fact that also the core has spun down means that there must be some
efficient torques accomplishing the angular momentum transport inside the Sun.
A likely candidate is the magnetic field.
It it indeed well known that only a weak poloidal field is needed to
brake the rotation of the radiative interior.

Helioseismology has indicated that the transition from latitudinal
differential rotation in the bulk of the convection zone to nearly rigid
rotation in the radiative interior is relatively sharp.
This transition region is called the tachocline.
The idea of a sharp transition region has problems of its own, because
viscous spreading would tend to smooth the transition with time.
The solution to this problem was thought to be related to the effect of
a mostly horizontal turbulence.
However, it can be argued that the rigidity of the radiative
interior is constantly maintained by the presence of a weak magnetic field
of about $1\G$; see R\"udiger \& Hollerbach (2004) for a recent monograph
covering also this aspect.

\subsection{Local helioseismology}

At larger values of $\ell$ the coherence time of the waves becomes
rather short and the modes are no longer global and take on a more local
character.
There are various techniques that use these modes to extract information
about local variations of sound speed and local flows.
The most popular method is the ring diagram technique.
For a detailed review see Gizon \& Birch (2005).
Among other things this method has demonstrated the presence of
converging flows around sunspots and a rather shallow temperature
subsurface structure.
However, a serious shortcoming of the present approach is the neglect
of magnetosonic and Alfv\'en waves.

\section{Solar activity cycle}

In the following we discuss some basic properties of the solar
magnetic field.
Its main feature is the 11 year cycle, as manifested in the (approximately)
eleven year variation of the sunspot number.
Sunspots are associated with sites of a strong magnetic field of
about $2$--$3\kG$ peak field strength.
Sunspots appear typically at about $\pm30^\circ$ latitude at the beginning
of each cycle, i.e.\ when the sunspot number begins to rise again.
During the course of the cycle, spots appear at progressively lower latitudes.
At the end of the cycle, sunspots appear at low latitudes of about $\pm4^\circ$.
Again, detailed references cannot be given here, but we refer to the
paper by Solanki et al.\ (2006) for a recent review.

\subsection{The butterfly diagram}

\begin{figure}[t!]\begin{center}
\includegraphics[width=\textwidth]{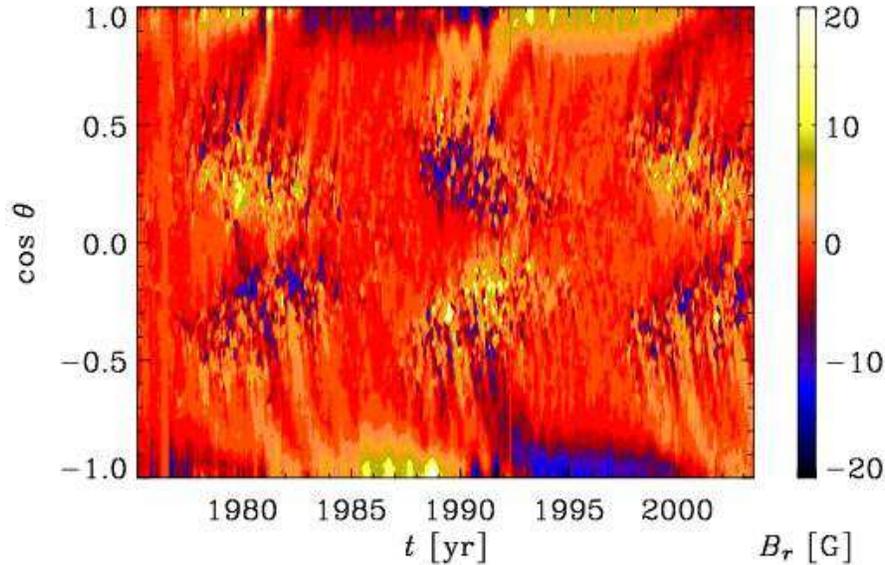}
\end{center}\caption[]{
Longitudinally averaged radial component of the observed solar
magnetic field as a function of cos(colatitude) and time.
Dark shades denote negative values and
light shades denote positive values.
Note the sign changes both in time and across the equator
(courtesy of R.~Knaack). 
}\label{knaack}\end{figure}

Although the detailed mechanism of their formation is still uncertain,
it seems that sunspots form when a certain threshold field is exceeded,
so they occur usually only below $\pm30^\circ$ latitude.
However, magnetic fields can still be detected at higher latitudes all
the way up to the poles using the Zeeman effect.
\FFig{knaack} shows, as a function of latitude and time,
the normal component of the azimuthally averaged surface field,
$\meanBB(R_\odot,\theta,t)$, where
\EQ
\meanBB(r,\theta,t)=\int_0^{2\pi}\BB\;{\dd\phi\over2\pi}.
\label{taver}
\EN
Such diagrams, which can also be produced for the mean number of sunspots
as function of time and latitude, are generally referred to as
butterfly diagrams.

Although the field strength in sunspots is about $2\kG$, when the field is
averaged in longitude only a small net field of about $\pm20\G$ remains.
Near the poles the magnetic field is more clearly defined because it
fluctuates less strongly in time near the poles than at lower latitudes.
A characteristic feature is that
the polar field changes sign shortly after each sunspot maximum.

At intermediate latitudes $|\cos\theta|=0.5...0.7$, corresponding to
a latitude, $90^\circ-\theta$, of $\pm(30^\circ...45^\circ)$, there are
characteristic streaks of magnetic activity that seem to move poleward
over a short time ($\sim1...2\yr$).
These streaks are rather suggestive of systematic advection by
poleward meridional circulation near the surface.
This indicates that the streaks are really just a consequence of the
remaining flux of decaying active regions being advected poleward from
lower latitudes.
Looking at a plot of the magnetic field at poorer resolution would
show what is known as the polar branch, whose presence has been found
previously through various other proxies (e.g.\ through the migration
of the line where prominences occur).
This has been reviewed in detail by Stix (1974).

In summary, the cyclic variation of the field
together with its latitudinal migration, and the alternating
orientation of bipolar magnetic regions
are the main systematic properties of the solar magnetic field.
In \Sec{DynamoTheory} we discuss theoretical approaches to the present
understanding of this phenomenon.

\subsection{Cyclic activity on other solar-like stars}

It should be noted that magnetic activity and activity cycles
are not unique to the Sun.
In fact, many stars with outer convection zones display magnetic activity,
as is evidenced by proxies such as the H and K line emission within the
Calcium absorption line.
This H and K line emission is caused by hot plasma that is confined
in the magnetic flux tubes in the coronae of these stars.
Among the solar-like stars of spectral type G and solar-like rotation,
many have cyclic magnetic
activity while others show time-independent magnetic activity that
is believed to be associated with the possibility that these stars are
in a grand minimum, such as the famous Maunder minimum.

\subsection{Grand minima}

Grand minima are recurrent states of global magnetic inactivity of
a star.
This behavior may be associated with the chaotic nature of the
underlying dynamo process.
For the Sun this behavior is evidenced through the record of the
Carbon 14 isotope concentrations in tree rings
as well as through the Beryllium 10 isotope concentrations
of ice core drillings from Greenland.
It is interesting to note that during the Maunder minimum between
1645 and 1700 the magnetic activity was not completely suppressed;
$^{10}\mbox{Be}$ still show cyclic variability, albeit with a somewhat
longer period of about 15 years.
Shortly after the Sun emerged from the Maunder minimum the sunspot
activity was confined only to the northern hemisphere.
This type of latitudinal asymmetry has been seen in some dynamo models that
display sporadically a mixture of modes that are symmetric and
antisymmetric about the equator.
For the Sun, some of the earlier grand minima have specific
names such as the Sp\"orer minimum (1420--1530),
the Wolf Minimum (1280--1340), and the Oort minimum (1010--1050).

Grand minima can be important for the Earth's climate.
For example the Maunder minimum is associated with the
`Little Ice Age' that occurred from 1560 to 1850.
During the 500 years before that the Sun was particularly active as is
evidenced by the high levels of $^{14}$C production: this was the
period when wine was made from grapes grown in England and when the
Vikings colonized Greenland.

By combining different proxies of solar activity, several
typical time scales can be identified, the Schwabe 11-year cycle,
the 88-year Gleissberg cycle, the 205-year De Vries cycle, and
the 2100 or 2300 year Hallstatt cycle.

\subsection{Active regions and active longitudes}

Active regions are complexes of magnetic activity out of which
sunspots, flares, coronal mass ejections, and several other phenomena
emerge with some preference over other regions.
These regions tend to be bipolar, i.e.\ they come in pairs of opposite
polarity and are roughly aligned with the east--west direction.

Over periods of up to half a year active regions appear preferentially
at the same longitude and follow a latitude-dependent rotation law.
An analysis of solar magnetograms show that
at the beginning of each cycle, when most of the activity
occurs at about $\pm30^\circ$ latitude, the rotation rate
of the active longitudes is less than
at the end of each cycle, when the typical latitude is
only $\pm4^\circ$ latitude.
There are various reports that these longitudes might be stable over
longer periods of time (so-called active longitudes), but this is still
very much a matter of debate.

The notion of field line anchoring is occasionally used in connection
with sunspot proper motions.
Long before the internal angular velocity was determined via
helioseismology, it was known that sunspots rotate faster than the
surface plasma.
Moreover, young sunspots rotate faster than old sunspots.
A common interpretation is that young sunspots are still anchored
at a greater depth than older ones, and that therefore the internal
angular velocity must decrease with height.
This provided also the basis for the classical mean field dynamo
theory  of the solar cycle according to which the radial angular
velocity gradient has to be negative.
This will be discussed in more detail in \Sec{DynamoTheory}.

With the advance of helioseismology, it has become clear that at low
latitudes the angular velocity decreases with radius throughout
the bulk of the convection zone.
A negative radial gradient exists only in the upper $30\Mm$
(sometimes referred to as the supergranulation layer).
Indeed, the very youngest sunspots have a rotation rate that is comparable
to or even slightly in excess of the fastest angular velocity seen with
helioseismology anywhere in the Sun (i.e.\ at $r/R_\odot\approx0.95$).

\subsection{Torsional oscillations}
\label{TorsionalOscillations}

At the solar surface the angular velocity varies with the 11 year cycle.
In other words, $\meanO$ at the surface (at $r=R_\odot$) is not only
a function of colatitude $\theta$, but also of time.
The pattern of $\meanO(R_\odot,\theta,t)$ shows an equatorward
migration, similar to the butterfly diagram of the mean poloidal
magnetic field in \Fig{knaack}.
Helioseismology has now established that this pattern extends at least
half way into the convection zone.
At the bottom of the convection zone the 11 year variation is not (yet?)
observed, but there is possibly a 1.3 year modulation of the local
angular velocity, although this is still unclear and debated (see the
review by Thompson et al.\ 2003).
In recent years this 1.3 year modulation has gone away, but it has been
speculated that the presence of a modulation may depend on the phase in
the cycle.

The 11 year cyclic modulation is known as torsional oscillation, but model
calculations demonstrate that these oscillations can be understood
as a direct response to the varying magnetic field.
The amplitude of the torsional oscillations is about 8\%, suggesting that
magnetic effects must be moderate and the fields of sub-equipartition
strength.

\section{Dynamo theory}
\label{DynamoTheory}

Given that the magnetic decay times in astrophysical plasmas
are generally very long, there have been a number of attempts in the literature
to explain the Sun's magnetic field in terms of a primordial, frozen-in field.
Such approaches tend to be rather sketchy when it comes to
predicting any quantitative details that can be tested.
Dynamo theory, on the other hand, provides a self-consistent framework of
magnetic field generation in general that can be tested against direct
simulations.
Owing to the turbulent nature of the flows, such dynamos are generally
referred to as ``turbulent dynamos''.
Unfortunately, early simulations did not reproduce the solar behavior
very well.
The reason for this may simply be that, for example, the resolution was
insufficient to capture important details.
The failure to explain the observations has led to a number of ad hoc
assumptions and modifications that are not satisfactory.
At the same time, dynamo theory itself has experienced some important
extensions that followed from trying to explain a long standing
mismatch between simulations and theory, even under rather idealized
conditions such as forced turbulence in a periodic domain.
In this section we can only outline the basic aspects of dynamo theory.
For a more extensive review, especially of the recent developments, see
Ossendrijver (2003) and Brandenburg \& Subramanian (2005).

\subsection{The induction equation}

At the heart of dynamo theory is the induction equation, which is
just the Faraday equation together with Ohm's law, i.e.\
\EQ
{\partial\BB\over\partial t}=-\nab\times\EE\quad\mbox{and}\quad
\JJ=\sigma\left(\EE+\UU\times\BB\right),
\EN
respectively.
The initial conditions furthermore must obey $\nab\cdot\BB=0$.
Eliminating $\EE$ yields
\EQ
{\partial\BB\over\partial t}
=\nab\times\left(\UU\times\BB-\JJ/\sigma\right).
\label{InductionEqn}
\EN
Then, using Ampere's law (ignoring the Faraday displacement current),
$\JJ=\nab\times\BB/\mu_0$,
where $\mu_0$ is the vacuum permeability, one obtains the induction
equation in a form that reveals the diffusive nature of the last term
as $...+\eta\nabla^2\BB$, where $\eta=(\sigma\mu_0)^{-1}$.

A complete theory of magnetic field evolution must include also the
momentum equation, because the magnetic field will react back on the
velocity field through the Lorentz force, $\JJ\times\BB$, so
\EQ
\rho{\DD\UU\over\DD t}=-\nab p+\JJ\times\BB+\FF,
\label{UU}
\EN
together with the continuity equation,
$\partial\rho/\partial t=-\nab\cdot(\rho\UU)$.
In \Eq{UU}, $\FF$ subsumes a range of possible additional forces
such as viscous and gravitational forces, as well as possibly
Coriolis and centrifugal forces.

To study the dynamo problem, the complete set of equations
is often solved using fully
three-dimensional simulations both in Cartesian and
in spherical geometries.
Especially in early papers, the continuity equation has been replaced by
the incompressibility condition, $\nab\cdot\UU=0$, or by the
anelastic approximation, $\nab\cdot(\rho\UU)=0$.
In both cases, $\rho$ no longer obeys an explicitly time-dependent
equation, and yet $\rho$ can of course change via the equation of state
(pressure and temperature are still changing).
These approximations are technically similar to that of neglecting the
Faraday displacement current.

As long as the magnetic field is weak, i.e.\ $\BB^2/\mu_0\ll\rho\UU^2$
at all scales and all locations, it may be permissible to assume $\UU$
as given and to solve only the induction equation for $\BB$.
This is called the kinematic dynamo problem.

Meanwhile some types of dynamos have been verified in experiments.
One is the Ponomarenko-like dynamo that consists of a swirling flow
surrounded by a nonrotating counterflow (Gailitis et al.\ 2001).
The flow is driven by propellers and leads to self-excited dynamo action
when the propellers exceed about 1800 revolutions per minute ($30\Hz$),
producing peak fields of up to $1\kG$.
Another experiment consists of an array of 52 connected tubes with
an internal winding structure through which liquid sodium is pumped,
making the flow strongly helical with nearly uniform kinetic helicity
density within the dynamo module containing the pipes
(Stieglitz \& M\"uller 2001).
Such a flow is particularly interesting because it allows meaningful
averages to be taken, making this problem amenable to a mean field
treatment.
The mean field approach is important in solar physics
and will be discussed in \Sec{MeanFieldTheory}.
First, however, we discuss the case where no mean field is produced
and only a small scale field may be generated.

\subsection{Small scale dynamo action}

There is an important distinction between small scale and
large scale turbulent dynamos.
This is mainly a distinction by the typical scale of the field.
Both types of dynamos have in general a turbulent component,
but large scale dynamos have an additional component
on a scale larger than the typical scale of the turbulence.
Physically, this can be caused by the effects of anisotropies,
helicity, and/or shear.
These large scale dynamos are amenable to mean field modeling (see below).
On the other hand, small scale dynamo action is possible under
fully isotropic conditions.
This process has been studied both analytically and numerically;
see Brandenburg \& Subramanian (2005) for a review.
Indeed, small scale dynamos tend to be quite prominent in
simulations, perhaps more so than what is realistic.
This may be a consequence of having used unrealistically large values
of the magnetic Prandtl number, as will be discussed in the following.

\begin{figure}[t!]\begin{center}
\includegraphics[width=.85\textwidth]{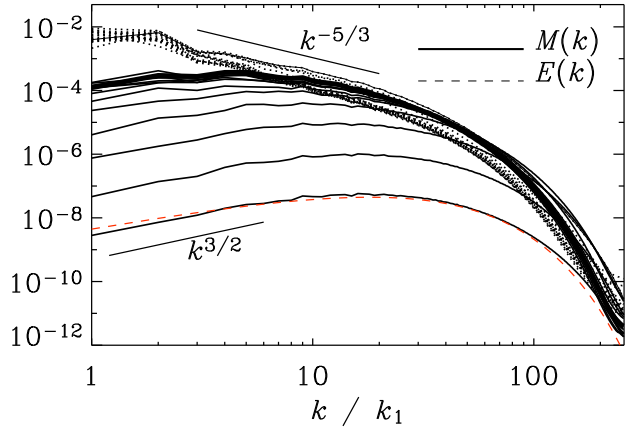}
\end{center}\caption[]{
Magnetic and kinetic energy spectra from a nonhelical turbulence
simulation with $P_{\rm m}=1$.
The kinetic energy is indicated as a dashed line (except for the first
time displayed where it is shown as a thin solid line).
At early times the magnetic energy spectrum follows the $k^{3/2}$
Kazantsev (1968) law (the dashed line gives the fit to the analytic spectrum),
while the kinetic energy shows a short $k^{-5/3}$ range.
The Reynolds number is $u_{\rm rms}/(\nu k_{\rm f})\approx600$
and $512^3$ meshpoints were used.
The time difference between the spectra is about
$14\,(k_{\rm f}u_{\rm rms})^{-1}$.
[Adapted from Brandenburg \& Subramanian (2005).]
}\label{pspec_nohel512d2}\end{figure}

The strength of the small scale dynamo depends significantly on the
value of the magnetic Prandtl number $\mbox{Pr}_{\rm M}\equiv\nu/\eta$,
i.e.\ the ratio of the viscosity, $\nu$, to the magnetic diffusivity $\eta$.
In the Sun, $\mbox{Pr}_{\rm M}$ varies between $10^{-7}$ and $10^{-4}$
between the top and the bottom of the convection zone, but it is always well
below unity.
In this case the Kolmogorov cutoff scale of the kinetic energy spectrum
of the turbulence is much smaller than the resistive cutoff scale of the
magnetic energy spectrum.
Therefore, at the resistive scale where the small scale dynamo would
operate fastest,
the velocity is still in its inertial range where
the spatial variation of the velocity is much more pronounced than
it would be near the Kolmogorov scale, relevant for a
magnetic Prandtl number of order unity.
This tends to inhibit small scale dynamo action.
In many simulations $\mbox{Pr}_{\rm M}$ is close to unity, because otherwise
the magnetic Reynolds number would be too small for the dynamo to be excited.
As a consequence, the production of small scale field may be exaggerated
in simulations.
It is therefore possible that in the Sun small scale dynamo action
is less important, and that large scale dynamo action is by comparison much
more prominent, than found in simulations.
An example may be the simulations of Brun et al.\ (2004),
which are currently the highest resolution turbulence simulations of
solar-like convection in spherical shell geometry.
Here the magnetic field is indeed mostly of small scale.

In mean field models only the large scale field is modeled.
This large scale field is governed both by
turbulent magnetic diffusion as well as non-diffusive contributions
such as the famous $\alpha$ effect.
As will be explained in the next section, this means that the mean
electromotive force has a component parallel to the mean field,
so it has a term of the form $\alpha\meanBB$;
see Brandenburg \& Subramanian (2005) for a recent review.
However, once a large scale field is present, the turbulent
motions (which are always present) will wind up and mix the large scale
field and will hence also produce a small scale field.
This does {\it not} represent small scale dynamo action, even though
there is a small scale field; if the large scale field is absent,
the small scale field disappears.

Let us emphasize again that the Sun does possess a large scale field,
with spatio-temporal order, as is evidenced by \Fig{knaack}.
This automatically implies a small scale field.
In addition, there may be small scale dynamo action occurring locally
in the near-surface layers where the Coriolis force is comparatively
weak, but this depends on whether or not small scale dynamo action
is inhibited by a small value of the magnetic Prandtl number.

\subsection{Mean field theory}
\label{MeanFieldTheory}

The mean field approach allows the complicated three-dimensional
dynamics to be treated in a statistical manner.
The averaged equations are then only two-dimensional.
In some cases, e.g.\ in Cartesian geometry, it can be useful to define
two-dimensional averages, so that the resulting mean field equations
are only one-dimensional.
In the following we describe the essential features of this approach.
By averaging the induction equation \eq{InductionEqn},
e.g.\ according to the toroidal averaging procedure, we obtain
\EQ
{\partial\meanBB\over\partial t}=\nab\times\left(
\meanUU\times\meanBB+\meanEMF-\eta\mu_0\meanJJ\right),
\label{InductionEqnMean}
\EN
where $\meanEMF=\overline{\uu\times\bb}$ is the mean electromotive
force from the small scale magnetic and velocity fields, with
$\uu=\UU-\meanUU$ and $\bb=\BB-\meanBB$ being the fluctuations,
i.e.\ the deviations from the corresponding averages.

There are two quite different approaches to calculating $\meanEMF$
and its dependence on $\meanBB$.
The first order smoothing approximation uses just the linearized
evolution equation for $\bb$, while the tau approximation uses also
the linearized momentum equation together with a closure hypothesis
for the higher order triple correlation terms.
For references and historical aspects we refer to the review by
Brandenburg \& Subramanian (2005).
Both approaches predict the presence of terms of the form
\EQ
\meanemfs_i=
\alpha_{ip}\meanB_p+\eta_{ipl}\meanB_{p,l},
\label{ubdot}
\EN
where a comma denotes partial differentiation.
The tau approximation gives
\EQ
\alpha_{ip}=-\tau\epsilon_{ijk}\overline{u_ku_{j,p}}
+\tau\epsilon_{ijk}\overline{b_kb_{j,p}}/\rho_0,
\label{alpipM}
\EN
where $\tau$ is the correlation time.
However, within the first order smoothing approximation the magnetic
term in $\alpha_{ip}$ is absent.
In order to illuminate the meaning of these tensors, it is useful
to make the assumption of isotropy,
$\tilde\alpha_{ip}=\tilde\alpha\delta_{ip}$ and
$\tilde\eta_{ipl}=\tilde\eta_{\rm t}\epsilon_{ipl}$.
This yields
\EQ
\tilde\alpha=-\onethird\left(\overline{\oo\cdot\uu}
                            -\overline{\jj\cdot\bb}/\rho_0\right),
\quad\tilde\eta_{\rm t}=\onethird\overline{\uu^2},
\EN
where $\oo=\nab\times\uu$ is the small scale vorticity
and $\jj=\nab\times\bb/\mu_0$ is the small scale current density.
Thus, $\tilde\alpha$ is proportional to the residual helicity, i.e.\
the difference between kinetic and current helicities, and $\eta_{\rm t}$
is proportional to the mean square velocity.

Using a closure assumption for the triple correlations we have,
under the assumption of isotropy, the important result
\EQ
\alpha=-\onethird\tau\left(\overline{\oo\cdot\uu}
                          -\overline{\jj\cdot\bb}/\rho_0\right),
\quad\eta_{\rm t}=\onethird\tau\overline{\uu^2}.
\label{ResidualHelEta}
\EN
The electromotive force takes then the form
\EQ
\meanEMF=\alpha\meanBB-\eta_{\rm t}\mu_0\meanJJ.
\label{alpBetaJ}
\EN
This equation shows that the electromotive force does indeed have a
component in the direction of the mean field (with coefficient $\alpha$).
The $\eta_{\rm t}$ term corresponds to a contribution of the
electromotive force that is formally similar to the microscopic
diffusion term, $\eta\mu_0\meanJJ$, in \Eq{InductionEqnMean}.
Therefore one speaks also of the total magnetic diffusivity,
$\eta_{\rm T}=\eta+\eta_{\rm t}$.
The presence of the $\alpha$ term, on the other hand, has no
correspondence to the non-turbulent case, and it is this term
that invalidates Cowling's anti-dynamo theorem for mean fields.
Indeed, there are simple self-excited (exponentially growing)
solutions already in a one-dimensional model (see below).

\EEq{ResidualHelEta} shows that the presence of an $\alpha$ effect
is closely linked to the presence of kinetic and/or current helicity,
while turbulent magnetic diffusion is always present when there is a
small scale turbulent velocity field.
This shows immediately that just increasing the turbulence (without also
increasing the helicity) tends to {\it diminish} turbulent mean field
dynamo action, rather than enhancing it, as one might have thought.

The formalism discussed above does not address the production of kinetic
helicity in the Sun.
This can be calculated perturbatively by considering the effects of
vertical density and turbulent intensity stratification and rotation.
At lowest order one finds
\EQ
\alpha_{\phi\phi}=-{\textstyle{16\over15}}\tau^2u_{\rm rms}^2
\Omega\cdot\nab\ln(\rho u_{\rm rms})+...
\label{KrausesFormula}
\EN
for the first term in \Eq{alpipM}.
For details we refer to the reviews by
R\"udiger \& Hollerbach (2004) and Brandenburg \& Subramanian (2005).
The magnetic contribution to the $\alpha$ effect proportional to
$\overline{\jj\cdot\bb}$ (in the isotropic case) still needs to be added
to the right hand side of \Eq{KrausesFormula}.
This $\overline{\jj\cdot\bb}$ contribution is mainly the result of the
dynamo itself, which tends to built up small scale current helicity
along with the large scale magnetic field.
Thus, the value of $\overline{\jj\cdot\bb}$ cannot be obtained
independently of the actual solution to the dynamo problem.

\begin{figure}[t!]\begin{center}
\includegraphics[width=.95\textwidth]{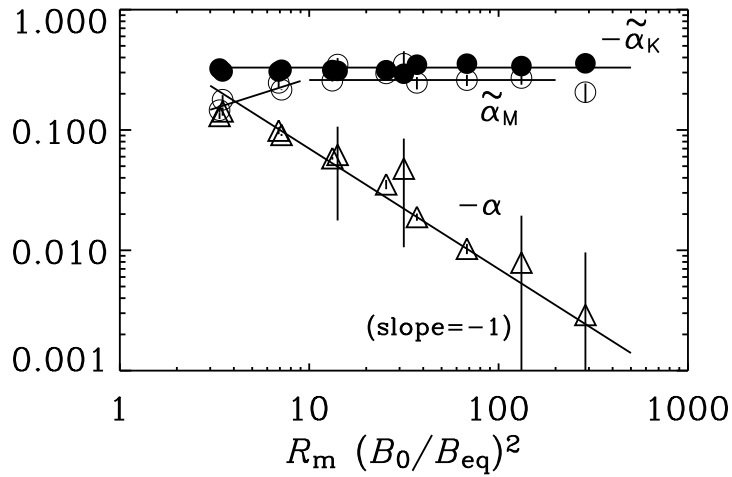}
\end{center}\caption[]{
$R_{\rm m}$ dependence of the normalized $\alpha$
compared with dependence of $\tilde{\alpha}_{\rm K}^{\rm(k)}$
for $B_0/B_{\rm eq}$ around unity.
Vertical bars give error estimates.
The vertical bars on the data points give estimates of the
error (see text).
[Adapted from Brandenburg \& Subramanian (2005).]
}\label{palp_vs_Rm_solint}\end{figure}

\subsection{Numerical determination of $\alpha$}

A simple way of determining $\alpha$ numerically is by imposing a constant
field of strength $\BB_0$ over a domain of simulated turbulence.
Since the mean field is constant, i.e.\ $\meanBB=\BB_0$, the mean current
density vanishes in \Eq{alpBetaJ},
so $\alpha$ can directly be determined by measuring
the electromotive force, $\overline{\uu\times\bb}$, in the direction
of the imposed field, and dividing one by the other.
In other words, $\alpha=\overline{\uu\times\bb}\cdot\BB_0/B_0^2$.
The values of $\alpha$ collapse onto a single line.
Looking at \Eq{ResidualHelEta}, such a decline of $\alpha$ can only
come about if either $\tau$ or $\overline{\oo\cdot\uu}$ decrease with $B_0$,
or, alternatively, if $\overline{\oo\cdot\uu}$ and $\overline{\jj\cdot\bb}$
approach each other.
It is quite clear from the data that neither $\overline{\oo\cdot\uu}$
nor $\tau$ decrease and that instead there is,
at least for small values of $R_{\rm m}(B_0/B_{\rm eq})^2$, a tendency
for $\overline{\oo\cdot\uu}$ and $\overline{\jj\cdot\bb}$ to approach
one another.

The ``catastrophic'' decrease of $\alpha$ with decreasing $\eta$ is
directly a consequence of magnetic helicity conservation in a closed
or periodic domain, but this can be alleviated in the presence of helicity
fluxes out of the domain.
We return to this discussion in \Sec{Nonlinear} when we consider
the consequences for the nonlinear saturation of the dynamo effect.

\subsection{Other effects}

There are a number of other effects that contribute to the algebraic
relationship between the electromotive force and the mean field.
One is a pumping effect associated with the antisymmetric components
of the $\alpha$ tensor,
\EQ
\alpha_{ij}^{\rm(A)}=\half(\alpha_{ij}+\alpha_{ji})\equiv
-\half\epsilon_{ijk}\gamma_k\quad\mbox{(pumping)},
\EN
where $\gamma_k$ is the pumping velocity.
This name is motivated by the fact that the term
$\alpha_{ij}^{\rm(A)}\meanB_j$ can also be written as
$(\ggamma\times\meanBB)_i$.
This shows that the vector $\ggamma$
plays the role of an effective advection velocity.

The pumping effect is sometimes called turbulent diamagnetism.
This has to do with a remarkable relation between pumping velocity and
turbulent magnetic diffusion,
\EQ
\ggamma=-\half\nab\eta_{\rm t}.
\EN
Calculating the contribution to the electromotive force from this term
together with the turbulent diffusion term gives
\EQ
\meanEMF=...-\half\nab\eta_{\rm t}\times\meanBB-\eta_{\rm t}\nab\times\meanBB
=...-{1\over\sigma_{\rm t}}\nab\times\left({\meanBB/\mu_{\rm t}}\right),
\EN
where
\EQ
\sigma_{\rm t}=\sigma(\eta_{\rm t}/\eta)^{-1/2}
\quad\mbox{and}\quad
\mu_{\rm t}=\mu_0(\eta_{\rm t}/\eta)^{-1/2}
\EN
are turbulent conductivity and turbulent permeability, respectively.
(The normalization with the microscopic values of $\sigma$ and $\mu_0$
is done in order for the turbulent values of $\sigma_{\rm t}$ and
$\mu_{\rm t}$ to have correct dimensions.)

Another potentially important term is an effect of the form
$\ddelta\times\meanJJ$, which has long been known to be able to produce
dynamo action if its components are of the appropriate sign relative to
the orientation of shear.
It is clear that $\ddelta$ must be an axial vector, and both the local
angular velocity, $\Omega$, as well as the vorticity of the mean flow,
$\meanWW=\nab\times\meanUU$ are known to contribute.
Dynamo action is only possible when $\ddelta$ and $\meanWW$ are antiparallel.
It is still not quite clear from turbulence calculations 
whether the orientation of the vector $\ddelta$
relative to the shear is appropriate for dynamo action in the convection zone.

\section{Models of the solar cycle}
\label{ModelsSolarCycle}

\subsection{One-dimensional models}

It has long been known that an $\alpha$ effect combined with differential
rotation can cause oscillatory propagating solutions.
In order to appreciate the possibility of oscillatory self-excited
solutions, let us consider one-dimensional solutions, allowing for
variations only in the $z$ direction, but field components still pointing
in the two directions.
Applied to the Sun, we may think of the $z$ direction being latitude
($=$~negative colatitude, $-\theta$), $x$ being radius, and $y$ being
longitude, so $(x,y,z)\longrightarrow(r,\phi,-\theta)$.
Let us consider a mean flow of the form $\meanUU=(0,Sx,0)$, i.e.\ the
flow has only a $y$ component that varies linearly in the $x$ direction.
We write the field in the form $\meanBB(z,t)=(-\meanA_y',\meanB_y,0)$,
where a prime denotes a $z$ derivative.
The corresponding dynamo equation can then be written as
\EQ
\dot{\meanA}_y=\alpha\meanB_y+(\eta+\eta_{\rm t})\meanA_y'',
\EN
\EQ
\dot{\meanB}_y=S\meanB_x+(\eta+\eta_{\rm t})\meanB_y'',
\EN
where we have neglected a term $(\alpha\meanB_x)'$ in comparison
with $S\meanB_x$ in the second equation.
(Here $\meanB_x=-\meanA_y'$ is the radial field.)

Solutions to these equations are frequently discussed in the literature
(e.g.\ Moffatt 1978, Brandenburg \& Subramanian 2005).
It is instructive to consider first solutions in an unbounded
domain, e.g.\ $0<z<L_z$, so the solutions are of the form
\EQ
\meanBB(z,t)=\mbox{Re}
\left[\hat{\BB}_\kk\exp\left(\ii kz+\lambda t\right)\right].
\EN
There are two physically meaningful solutions.
Both have an oscillatory component, but one of them can also have an
exponentially growing component such that real and imaginary parts of
$\lambda$ are given by
\EQ
\mbox{Re}\lambda=-\eta_{\rm T}k^2+\left|\half\alpha Sk\right|^{1/2},
\EN
\EQ
\mbox{Im}\lambda\equiv-\omega_{\rm cyc}=\left|\half\alpha Sk\right|^{1/2}.
\EN
The solutions are oscillatory with the cycle period $\omega_{\rm cyc}$.
This shows that, in the approximation where the $(\alpha\meanB_x)'$ term is
neglected (valid when $Sk\gg\alpha$), the mean field dynamo is excited when
the dynamo number,
\EQ
D=|\half\alpha Sk|^{1/2}/(\eta_{\rm T}k^2),
\EN
exceeds a critical value that is in this simple model $D_{\rm crit}=1$.

A number of important conclusions can be drawn based on this simple model.
(i) The cycle frequency is proportional to $\sqrt{\alpha S}$, but becomes
equal to $\eta_{\rm T}k^2$ in the marginal or nonlinearly saturated cases.
(ii) There are dynamo waves with a pattern speed proportional to
$\eta_{\rm T}k$ propagating along contours of constant shear.
For example, for radial angular velocity contours with angular
velocity decreasing outwards, and for a positive $\alpha$ in the
northern hemisphere, the propagation is equatorward.
If the sign of either $S$ or $\alpha$ is reversed,
the propagation direction is reversed too.

For more realistic applications to the Sun one must solve the mean field
dynamo equations in at least two dimensions over a spherical domain with
appropriate profiles for $\alpha$, $\eta_{\rm T}$, and $\Omega$.
In the following we discuss four different dynamo scenarios that have
been studied over the years.

If the flow is assumed given, no feedback via the Lorentz force is allowed,
so the dynamo equations are linear and the magnetic energy would eventually
grow beyond all bounds.
In reality, the magnetic field will affect the flow and hence
$\meanUU$, as well as $\alpha$, $\eta_{\rm t}$, and other turbulent
transport coefficients will be affected.
We will postpone the discussion of the nonlinear behavior to \Sec{Nonlinear}.

\subsection{Different solar dynamo scenarios}

\begin{figure}[t!]\begin{center}
\includegraphics[width=\textwidth]{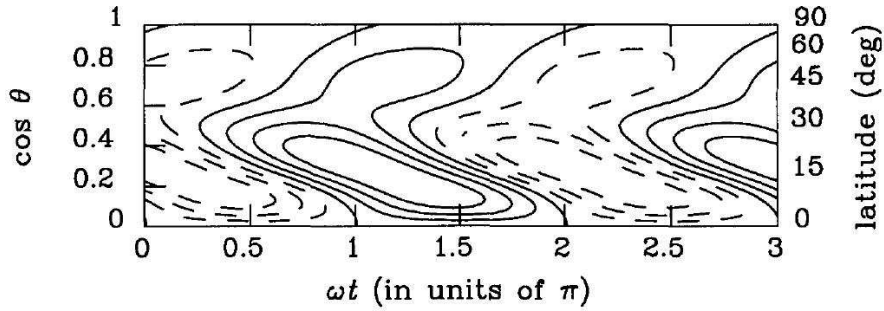}
\end{center}\caption[]{
Butterfly diagram of $B_\phi$ taken at reference depth $r=0.85R_\odot$.
[Adapted from Brandenburg \& Tuominen (1988).]
}\label{distributed}\end{figure}

A traditional and also quite natural approach is to calculate the profiles
for $\alpha$ and $\eta_{\rm t}$ using the results from mean field theory
such as \Eq{KrausesFormula}
and to take the profiles for the rms velocity and the correlation time,
$\tau=\ell/u_{\rm rms}$, from stellar mixing length models using
$\ell=\alpha_{\rm mix}H_p$ for the mixing length, where $H_p$ is
the pressure scale height.
For $\Omega(r,\theta)$ one often uses results from helioseismology.
In \Fig{distributed} we reproduce the results of an early paper where
the $\Omega(r,\theta)$ profile was synthesized from a collection of
different helioseismology results then available.
The $\alpha$ and  $\eta_{\rm t}$ profiles, as well as profiles describing
some other effects (such as pumping and $\OOmega\times\meanJJ$ effects)
where taken from a solar mixing length model.
In this model an equatorward migration is achieved in a limited
range in radius where $\partial\Omega/\partial r<0$.
In this model this is around $r=0.8R_\odot$.
Note also that in this case $\meanB_r$ and $\meanB_\phi$ are approximately
in antiphase, as is also seen in the Sun.

Distributed dynamos have been criticized on the grounds that magnetic
buoyancy will rapidly remove the magnetic field from the convection zone.
Since then, helioseismology has shown that the radial $\Omega$ gradient
is virtually zero in the bulk of the convection zone and only at the
bottom is there a finite gradient, but it is positive at latitudes below
$\pm30^\circ$.
This may still yield an equatorward migration in the butterfly diagram,
because \Eq{KrausesFormula} would predict that at the bottom of the
convection zone, where the magnitude of the positive $\nabla_r\ln u_{\rm rms}$
gradient exceeds that of the negative $\nabla_r\ln\rho$ gradient.
This changes the sign of $\alpha$, and makes it negative near the
bottom of the convection zone in the northern hemisphere.
This led to the idea of the overshoot dynamo that is believed to operate
only in a thin layer at or just below the convection zone proper.
Such dynamos have been considered by a number of different groups.

\begin{figure}[t!]\begin{center}
\includegraphics[width=\textwidth]{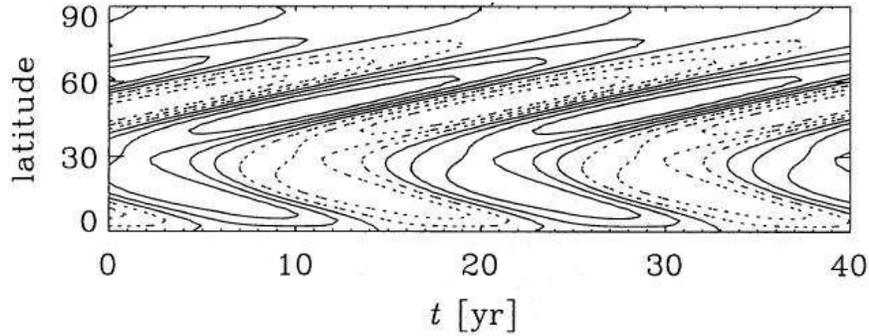}
\end{center}\caption[]{
Butterfly diagram of $B_\phi$ evaluated at the bottom of the convection
zone at $r=0.7R_\odot$.
[Adapted from R\"udiger \& Brandenburg (1995).]
}\label{overshoot}\end{figure}

In \Fig{overshoot} we show the result of an overshoot dynamo calculation.
An important problem that emerges from such an approach is that when the
dynamo layer is too thin, the toroidal flux belts are too close to each
other in latitude.
This leads to the conclusion that the thickness
of the dynamo region should not be less than $35\Mm$.
At the bottom of the convection zone this corresponds to half a pressure
scale height.
However, this value is already rather
large and no longer supported by helioseismology,
which predicts the thickness of the overshoot layer to be about $7\Mm$
or less.

Another variant of this approach is the interface dynamo.
The main difference here is that $\alpha$ is assumed to operate in
the bulk of the convection zone, but it is still taken to be negative,
so as to give equatorward migration.
Also important is the sharp jump in $\eta_{\rm t}$ at the bottom of the
convection zone.
However, when the latitudinal variation of the angular velocity is
included, no satisfactory butterfly diagram is obtained.

\begin{figure}[t!]\begin{center}
\includegraphics[width=\textwidth]{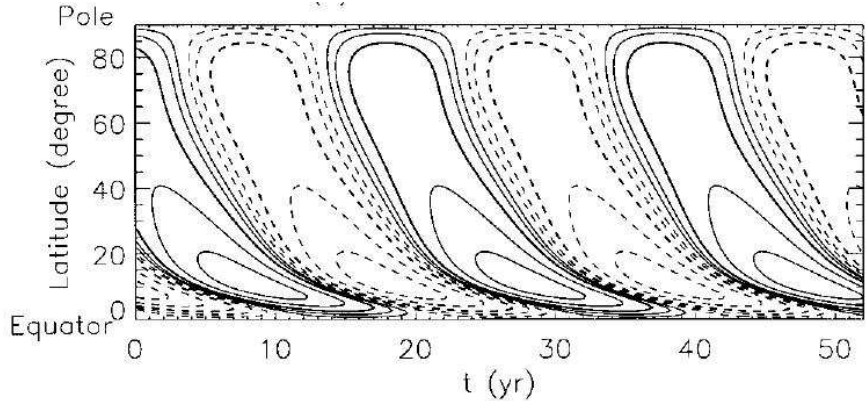}
\end{center}\caption[]{
Butterfly diagram of $B_\phi$ at $r=0.7R_\odot$.
The maximum circulation speed at the surface is $15\m\s^{-1}$
and the turbulent magnetic diffusivity is assumed to be
$\eta_{\rm t}=3\times10^{11}\cm^2\s^{-1}$.
[Courtesy of Dikpati \& Charbonneau (1999).]
}\label{FluxTransport}\end{figure}

A completely different class of dynamos are the flux transport dynamos
that are governed by the effect of meridional
circulation transporting surface flux to the poles and flux along the
tachocline toward the equator.
The $\alpha$ effect is now assumed positive, so in the absence of meridional
circulation the dynamo wave would propagate poleward.
However, under certain conditions, meridional circulation can actually
reverse the direction of propagation of the dynamo wave.
A calculation with a realistic solar angular velocity profile has been
presented by Dikpati \& Charbonneau (1999); see \Fig{FluxTransport}.
They establish a detailed scaling law for the dependence of the cycle
period on the circulation speed, the $\alpha$ effect (or source term),
and the turbulent magnetic diffusivity.
To a good approximation they find the cycle period to be
inversely proportional to the circulation speed.

With such a variety of different models and assumptions (most of them
ignoring what was previously derived for $\alpha(r,\theta)$,
$\eta_{\rm t}(r,\theta)$, and other transport effects), dynamo theory
has been perceived as rather arbitrary.
One reason for this level of arbitrariness that developed in modeling
the solar dynamo is that the effects of nonlinearity are not well understood.
This might affect the properties of the dynamo coefficients in the saturated
state making them quite different from those obtained in linear theory.
In the following we sketch briefly the tremendous developments
on nonlinear saturation that have occurred in the past few years.

\subsection{Nonlinear saturation}
\label{Nonlinear}

The effects of nonlinearity can be divided into macroscopic and
microscopic effects.
The former is simply the result of $\meanBB$ on $\meanUU$, as described
by the Lorentz force, $\meanJJ\times\meanBB$, in the mean field momentum
equation.
This effect is sometimes also referred to as the
Malkus--Proctor effect and has been incorporated to various degree
of sophistication in a number of models starting with incompressible
models in the context of the geodynamo and the solar dynamo.

The microscopic feedback can be subdivided into two different
contributions.
The effect of $\meanBB$ on the turbulent velocity
(conventional $\alpha$ quenching), and the more
direct effect of the small scale current helicity,
$\overline{\jj\cdot\bb}$ (or $\epsilon_{ijk}\overline{b_kb_{j,p}}$
in the anisotropic case), on the $\alpha$ effect or, more precisely,
on the electromotive force.
The latter can, under some conditions, lead to catastrophic $\alpha$
quenching; see Brandenburg \& Subramanian (2005) for a review of this
vast field of recent research.

The $\overline{\jj\cdot\bb}$ term cannot be implemented directly,
because it is necessary to have a theory for how $\overline{\jj\cdot\bb}$
depends on the mean field.
Under some idealized conditions (steady state, triply-periodic
boundary conditions) the answer can be obtained from the general evolution
equation for magnetic helicity, which reads
\EQ
{\partial\over\partial t}(\AAA\cdot\BB)+\nab\cdot\FFF_{\rm H}
=-2\eta\mu_0(\JJ\cdot\BB).
\EN
Here, $\AAA$ is the magnetic vector potential with $\BB=\nab\times\AAA$,
while $\AAA\cdot\BB$ is the magnetic helicity density, and
$\FFF_{\rm H}$ is its flux.
Magnetic helicity and its flux are gauge-dependent, i.e.\ they are not
invariant under the transformation $\AAA\to\AAA'=\AAA+\nab\Lambda$.
However, when averaging over a triply-periodic volume this ambiguity
disappears and $\bra{\AAA\cdot\BB}$ is gauge-invariant and obeys
\EQ
{\dd\over\dd t}\bra{\AAA\cdot\BB}=-2\eta\mu_0\bra{\JJ\cdot\BB}.
\label{dABperi}
\EN
(The spatial average of a divergence also vanishes for triply periodic
domains.)
We see that in the steady state, ${\rm d}/{\rm d}t=0$, so
$\bra{\JJ\cdot\BB}=0$.
Splitting into large scale and small scale contributions, we have
$\bra{\jj\cdot\bb}=-\bra{\meanJJ\cdot\meanBB}$.
This connects the small scale current helicity explicitly
with the properties of the large scale field.

The same procedure can still be applied in the unsteady case by
considering magnetic helicity evolution for the large scale and
small scale components, i.e.\ for $\bra{\meanAA\cdot\meanBB}$
and $\bra{\aaa\cdot\bb}$.
The evolution of $\bra{\meanAA\cdot\meanBB}$ follows
straightforwardly from the mean field equations, which shows that
there is continuous production of large scale magnetic helicity
given by $2\meanEMF\cdot\meanBB$.
In order not to produce any net magnetic helicity, as required by
\Eq{dABperi}, the evolution of $\bra{\aaa\cdot\bb}$ has the same term
but with the opposite sign.
Furthermore, under isotropic conditions, $\bra{\aaa\cdot\bb}$ is
proportional to $\bra{\jj\cdot\bb}$, which in turn is proportional
to the magnetic contribution to the $\alpha$ effect.
Finally, the restriction to triply periodic boundary conditions can
be relaxed (and hence a flux divergence can be permitted) if there is
sufficient scale separation, i.e.\ if the energy carrying scale of the
turbulence is clearly smaller than the domain size.

This then leads to an explicit evolution equation for the magnetic
$\alpha$ effect,
\EQ
{\partial\alpha_{\rm M}\over\partial t}+\nab\cdot\meanFF_\alpha
=-2\eta_{\rm t}k_{\rm f}^2\left({\meanEMF\cdot\meanBB\over B_{\rm eq}^2}
+{\alpha_{\rm M}\over R_{\rm m}}\right),
\EN
where $\alpha_{\rm M}=\onethird\tau\overline{\jj\cdot\bb}$ is the
magnetic $\alpha$ effect and $\meanFF_\alpha=\onethird\tau\meanFF_{\rm C}$,
where $\meanFF_{\rm C}\approx k_{\rm f}^2\meanFF_{\rm H}$ is the current
helicity flux.
This so-called dynamical $\alpha$ quenching equation is able to reproduce
the resistively slow saturation behavior,
found in simulations of helically driven turbulence.
In the steady state, this equation predicts for
$\alpha=\alpha_{\rm K}+\alpha_{\rm M}$
\EQ
\alpha={\alpha_{\rm K}+R_{\rm m}\left(\eta_{\rm t}\meanJJ\cdot\meanBB
+\nab\cdot\meanFF_\alpha\right)\over1+R_{\rm m}\meanBB^2/B_{\rm eq}^2}.
\EN
Note that in the special case of periodic domains, used in some
simulations where $\meanJJ=\bm{0}$ and $\nab\cdot\meanFF_\alpha=0$,
this equation
predicts catastrophic quenching, i.e.\ 
$\alpha=\alpha_{\rm K}/(1+R_{\rm m}\meanBB^2/B_{\rm eq}^2)$,
so $\alpha$ is suppressed relative
to its kinematic value $\alpha_{\rm K}$ in a strongly Reynolds number-dependent
fashion--as seen in \Fig{palp_vs_Rm_solint}.

In the case of open boundaries, there is a flux of magnetic helicity.
Under the two-scale hypothesis this can be defined in a gauge-invariant
manner.
Magnetic helicity fluxes provide a way to escape the otherwise
resistively limited saturation and catastrophic quenching.
Several simulations and mean field models have confirmed this.
Although dynamical quenching has already been applied to
solar dynamo models it remains to be seen
to what extent the previously discussed conclusions about
distributed versus overshoot layer dynamos are affected, and
what the role of meridional circulation is in such a model.

\subsection{Location of the dynamo}

It is generally believed that the magnetic field emergence in the
form of sunspots is deeply rooted and associated with strong toroidal
flux tubes of strength up to $100\kG$, as predicted by the so-called
thin flux tube models.
When parts of the flux tube become destabilized due to magnetic
buoyancy, it rises to the surface to form a sunspot pair.
However, there are some open questions: how are such coherent tubes
generated and what prevents them from breaking up during the ascent
over 20 pressure scale heights?
Alternatively, the usual mean field dynamo would actually predict magnetic
field generation distributed over the entire convection zone.
Sunspot formation would mainly be associated with local flux concentration
within regions of enhanced net flux.
This picture is appealing in many ways and has been discussed in
more detail in Brandenburg (2005).
However, although both pictures (deep rooted versus distributed dynamo)
have received some support from mean field modeling, there is still
no global turbulence simulations that reproduces the solar activity
cycle without questionable assumptions.

\section{Differential rotation}

It became clear from the discussion in \Sec{ModelsSolarCycle} that
differential rotation plays an important role in producing a large
scale magnetic field in the Sun.
It may also be important for the dynamo in disposing of its excess
small scale current helicity, as discussed in the previous section.
In this section we discuss the theoretical basis for explaining the
origin and properties of solar and stellar differential rotation.

\subsection{Mean field theory of differential rotation}

The origin of differential rotation has long been understood to be
a consequence of the anisotropy of convection.
It has long been clear that the
vertical exchange of momentum by convection should lead to a tendency
toward constant angular momentum in the radial direction,
i.e.\ $\meanO\varpi^2=\mbox{const}$, and hence the mean angular
velocity scales with radius like $\meanO(r)\sim r^{-2}$.
Here, $\varpi=r\sin\theta$ denotes the cylindrical radius
(i.e.\ the distance from the rotation axis).

The $r\phi$ component of the viscous stress tensor
contributes to the angular momentum equation,
\EQ
{\partial\over\partial t}\left(\rho\varpi^2\meanO\right)+
\nab\cdot\left[\rho\varpi\left(\meanUU\,\meanU_\phi+\overline{\uu u_\phi}
\right)\right]=0,
\label{AMeqn}
\EN
where $\overline{u_i u_j}=Q_{ij}$ are the components of the Reynolds tensor.
In spherical coordinates the full mean velocity vector is written as
$\meanUU=(\meanU_\varpi,\varpi\meanO,\meanU_z)$.

The early treatment in terms of an anisotropic viscosity tensor
was purely phenomenological.
A rigorous calculation of the Reynolds stresses shows that the
mean Reynolds stress tensor is described not only by diffusive
components that are proportional to the components of
the rate of strain tensor of the
mean flow, but that there are also non-diffusive components that
are directly proportional to the local angular velocity.
In particular the $r\phi$ and $\theta\phi$ components of the
Reynolds tensor are of interest for driving $r$ and $\theta$
gradients of $\meanU_\phi\equiv\varpi\meanO$.
Thus, for ordinary isotropic turbulent viscosity one has, using Cartesian
index notation,
\EQ
Q_{ij}=-\nu_{\rm t}\left(\meanU_{i,j}+\meanU_{j,i}\right)
-\zeta_{\rm t}\delta_{ij}\meanU_{k,k},
\EN
where $\zeta_{\rm t}$ is a turbulent bulk viscosity, and commas denote
partial differentiation.
This expression implies in particular that
\EQ
Q_{\theta\phi}=-\nu_{\rm t}\sin\theta{\partial\meanO\over\partial\theta}.
\EN
Note that for the Sun, where $\partial\meanO/\partial\theta>0$
in the northern hemisphere, this formula would predict that 
$Q_{\theta\phi}$ is negative in the northern hemisphere.
However, it was noted long ago from correlation measurements of
sunspot proper motions that
$Q_{\theta\phi}$ is in fact {\it positive} in the northern hemisphere.
The observed profile of $Q_{\theta\phi}$ is also known
as the Ward profile.
The observed positive sign was used to motivate
that there must be an additional term in the expression for $Q_{ij}$.
Using a closure approach, such as the
first order smoothing approximation that is often used
to calculate the $\alpha$ effect in dynamo theory, one can find
the coefficients in the expansion
\EQ
Q_{ij}=\Lambda_{ijk}\meanO_k-{\cal N}_{ijkl}\meanU_{k,l},
\EN
where $\Lambda_{ijk}$ describes
the so-called $\Lambda$ effect and ${\cal N}_{ijkl}$
is the turbulent viscosity tensor.
The viscosity tensor ${\cal N}_{ijkl}$ must in general be anisotropic.
When anisotropies are included, ${\cal N}_{ijkl}$ gets modified
(but it retains its overall diffusive properties),
and $\Lambda_{ijk}$ takes the form
\EQ
\Lambda_{ijk}\meanO_k=\pmatrix{
0 & 0 & V\sin\theta\cr
0 & 0 & H\cos\theta\cr
V\sin\theta & H\cos\theta & 0}\meanO,
\EN
where $V$ and $H$ are still functions of radius, latitude, and time;
$V$ is thought to be responsible for driving vertical differential rotation
($\partial\meanO/\partial r\neq0$) while $H$ is responsible for
latitudinal differential rotation ($\partial\meanO/\partial\theta\neq0$).

The first order smoothing approximation predicts the following
useful approximations for $V$ and $H$:
\EQ
V\approx\tau\left(\overline{u_\phi^2}-\overline{u_r^2}\right),
\EN
\EQ
H\approx\tau\left(\overline{u_\phi^2}-\overline{u_\theta^2}\right).
\EN
These expressions show that when the rms velocity in the radial direction
is larger than in the azimuthal direction we must expect $V<0$ and hence
$\partial\meanO/\partial r<0$.
In the Sun, this effect is responsible for the negative radial
shear near the surface where strong downdrafts may be responsible
for a comparatively large value of $\overline{u_r^2}$.
Likewise, when the rms velocity in the latitudinal direction is
larger than in the azimuthal direction we expect $H<0$ and hence
$\partial\meanO/\partial\theta<0$, so the equator would spin slower than
the poles.
This does not apply to the Sun, but it may be the case in some stars,
especially when the flows are dominated by large scale meridional
circulation.

\subsection{The $\Lambda$ effect from turbulence simulations}

Several of the relationships described above have been tested using
convection simulations, both in local Cartesian boxes located at different
latitudes as well as in global spherical shells.
Generally, the various simulations agree in that the sign of the horizontal
Reynolds stress is positive in the northern hemisphere and negative in
the southern, reproducing thus the Ward profile.
The simulations also show that the off-diagonal components
of the turbulent heat transport tensor are mostly
positive in the northern hemisphere, and negative in the southern hemisphere.
This agrees with the sign required if the baroclinic term is to produce
a tendency toward spoke-like angular velocity contours.
Simulations also reproduce the sudden drop of angular
velocity at the top of the convection zone.
This agrees with a predominantly negative sign of the vertical Reynolds
stress at a similar depth.
Furthermore, some of the more recent simulations show an unexpectedly sharp
increase of the horizontal Reynolds stress just near the equator (at
around $\pm5^\circ$ latitude), before changing sign right at the equator.
The significance of this result for the solar differential rotation
pattern is still unclear.

\subsection{Meridional flow and the baroclinic term}

According to the formalism described in the previous section,
a finite differential
rotation can be obtained by ignoring meridional flows and solving
\Eq{AMeqn} in isolation.
However, this would only be a poor approximation that becomes
quickly invalid when the angular velocity becomes large compared
with the turbulent viscous decay rate.
This is quantified by the Taylor number
\EQ
\mbox{Ta}=\left(2\meanO_0 R^2/\nu_{\rm t}\right)^2.
\EN
Using the first order smoothing expression from R\"udiger (1989),
$\nu_{\rm t}=(2/15)\,\tau u_{\rm rms}^2$, we have for
values typical for the Sun (see \Tab{SolarModel}), i.e.\
$\nu_{\rm t}\approx10^{12}$cm$^2$/s,
$\mbox{Ta}\approx10^9$.
This value of $\mbox{Ta}$ is rather large so that nonlinearities produce
strong deviations from linear theory.

\begin{figure}[t!]\begin{center}
\includegraphics[width=\textwidth]{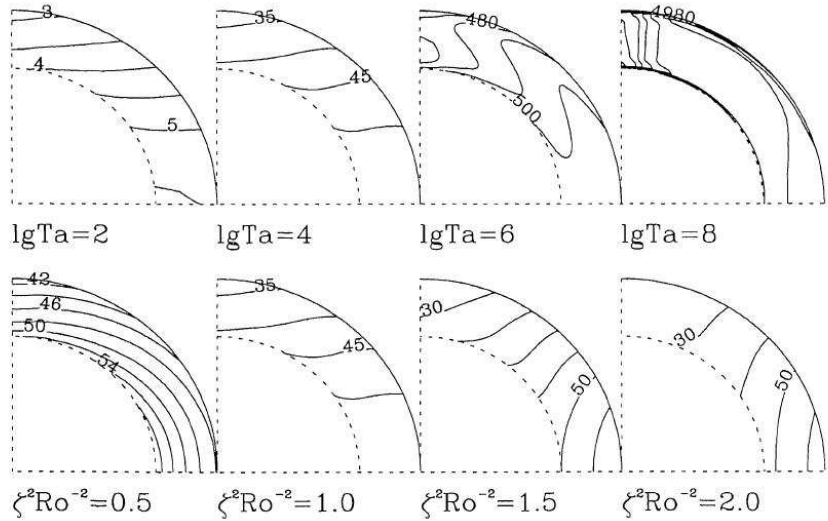}
\end{center}\caption[]{
Contours of constant $\meanO$ for different values of Taylor number (upper
panel) and different values of the inverse Rossby number, affecting the
relative importance of $H$ over $V$ (lower panel).
[Adapted from Brandenburg et al.\ (1990).]
}\label{OmegaTaylor}\end{figure}

As the value of $\mbox{Ta}$ is increased, the Coriolis force
increases, which then drives a meridional flow.
This meridional flow first increases with increasing values of $\mbox{Ta}$,
but then it reaches a maximum at $\mbox{Ta}\approx3\times10^5$, and later
declines with increasing values of $\mbox{Ta}$.
(The solar value is $\mbox{Ta}\approx3\times10^7$.)
This decline is because eventually the Coriolis force can no longer be
balanced against advection or diffusion terms.
This can best be seen by considering the curl of the momentum equation,
\EQ
{\partial\meanW_\phi\over\partial t}
+\varpi\meanUU\cdot\nab\left({\meanW_\phi\over\varpi}\right)
-\nu_{\rm t}\DD^2\!\meanW_\phi
=\varpi{\partial\meanO^2\over\partial z}
+\pp\cdot\overline{\nab T\times\nab S}.
\label{DwDt}
\EN
We recall that we consider here a nonrotating frame of reference,
so there is no Coriolis force.
Nevertheless, part of the inertial term takes a form that is
quite similar to the Coriolis term, but here $\meanO$ is a
function of position, while in the Coriolis term the angular
velocity would normally be a constant.

In the barotropic case one has $\nab T\parallel\nab S$ so there
is no baroclinic term, i.e.\ $\pp\cdot(\nab T\times\nab S)=0$.
So, if viscous and inertial terms are small, which is indeed the case
for rapid rotation, then $\partial\meanO^2/\partial z$ has to vanish,
so $\meanO$ would be constant along cylinders; see \Fig{OmegaTaylor}.
It is generally believed that the main reason for $\meanO$ not
having cylindrical contours in the Sun is connected with the
presence of the baroclinic term.
The presence of magnetic forces may also play a role, but unlike the
baroclinic term, magnetic forces tend to produce a rather variable
$\meanO$ patterns, often connected with rapid motions near the poles
where the inertia is lower.

Currently the highest resolution simulations of global convection in
spherical shells are those by Miesch et al.\ (2000).
These simulations show a great amount of detail and reproduce some
basic features of the Sun's differential rotation such as the
more rapidly spinning equator.
However, in low latitudes they show strongly cylindrical $\meanO$
contours that deviate markedly from the more spoke-like contours
inferred for the Sun using helioseismology.
These simulations also do not show the near-surface shear layer
where the rotation rate drops by over $20\nHz$ over the last
$30\Mm$ below the surface.

\begin{figure}[t!]\begin{center}
\includegraphics[width=.9\textwidth]{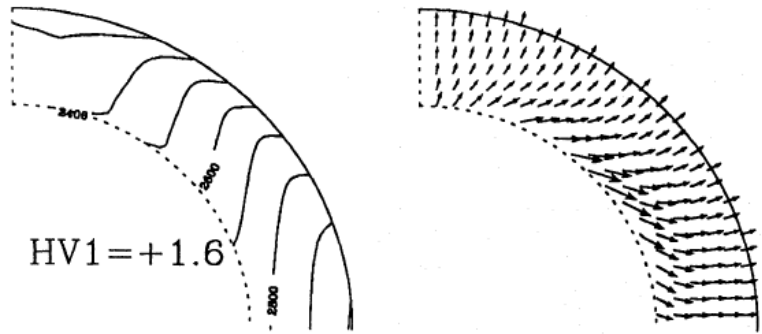}
\end{center}\caption[]{
Contours of angular velocity (left) and turbulent convective energy
flux (right) for a model with anisotropic heat transfer tensor.
[Adapted from Brandenburg et al.\ (1992).]
}\label{BMT92}\end{figure}

Mean field simulations using the $\Lambda$ effect
show surprisingly good agreement with the helioseismologically
inferred $\meanO$ pattern, and they are also beginning to address the
problem of the near-surface shear layer.
In these simulations it is indeed the baroclinic term that is
responsible for causing the departure from cylindrical contours.
This, in turn, is caused by an anisotropy of the turbulent heat
conductivity which causes a slight enhancement in temperature and
entropy at the poles.
In the bulk of the convection zone the entropy is nearly constant,
so the radial entropy variation is smallest compared with the radial
temperature variation.
It is therefore primarily the latitudinal entropy variation that determines
the baroclinic term, with
\EQ
\varpi{\partial\meanO^2\over\partial z}\approx-\pp\cdot
\overline{\nab T\times\nab S}
\approx-{1\over r}
{\partial\overline{T}\over\partial r}
{\partial\overline{S}\over\partial\theta}<0.
\EN
The inequality shows that
negative values of $\partial\meanO^2/\partial z$
require that the pole is slightly warmer than the equator
($\partial\overline{S}/\partial\theta<0$).
However, this effect is so weak that it cannot at present be observed.
Allowing for these conditions in a simulation may require particular care
in the treatment of the outer boundary conditions.
In \Fig{BMT92} we show the plots of angular velocity contours and convective
energy transport in a model with anisotropic turbulent conductivity tensor,
$\chi_{ij}$.
Given that the flux, $\FF$, is proportional
to $-\chi_{ij}\nabla_j\overline{S}$, a negative
$\partial\overline{S}/\partial\theta$ can be produced from a positive
$F_r$ with a positive value of $\chi_{r\theta}$.

In the discussion above we ignored in the last step a
possible correlation between entropy and temperature fluctuations,
i.e.\ a contribution from the term $\overline{\nab T'\times\nab S'}$
where primes denote fluctuations.
Such correlations, if of suitable sign,
might provide yet a further explanation
for a non-zero value of $\partial\meanO^2/\partial z$.

\subsection{Near-surface shear layer}

\begin{figure}[t!]\begin{center}
\includegraphics[width=.99\textwidth]{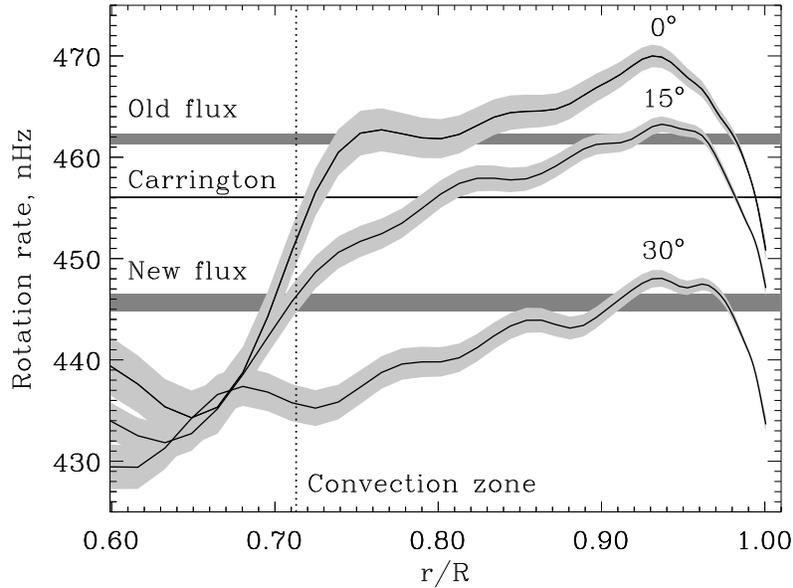}
\end{center}\caption[]{
Radial profiles of the internal solar rotation rate, as inferred from
helioseismology (sidereal, i.e.\ in a fixed frame).
The rotation rate of active zones at the beginning of the cycle
(at $\approx30^\circ$ latitude) and near the end (at $\approx4^\circ$)
is indicated by horizontal bars, which intersect the profiles of
rotation rate at $r/R_\odot\approx0.97$.
For orientation, the conventionally defined Carrington rotation period of
27.3~days (synodic value, corresponding to $424\nHz$) has been translated
to the sidereal value of $456\nHz$.
Courtesy of Benevolenskaya et al.\ (1999).
}\label{bene99}\end{figure}

The first results of helioseismology indicated significantly higher
angular velocities in the sub-surface than what is seen at the surface
using Doppler measurements.
This apparent conflict is now resolved in that helioseismological
inversions of the data from the SOHO spacecraft show a sharp negative
gradient, connecting the observed surface values smoothly with the
local maximum of the angular velocity at about $35\Mm$ depth;
see \Fig{bene99}.

\begin{figure}[t!]\begin{center}
\includegraphics[width=.99\textwidth]{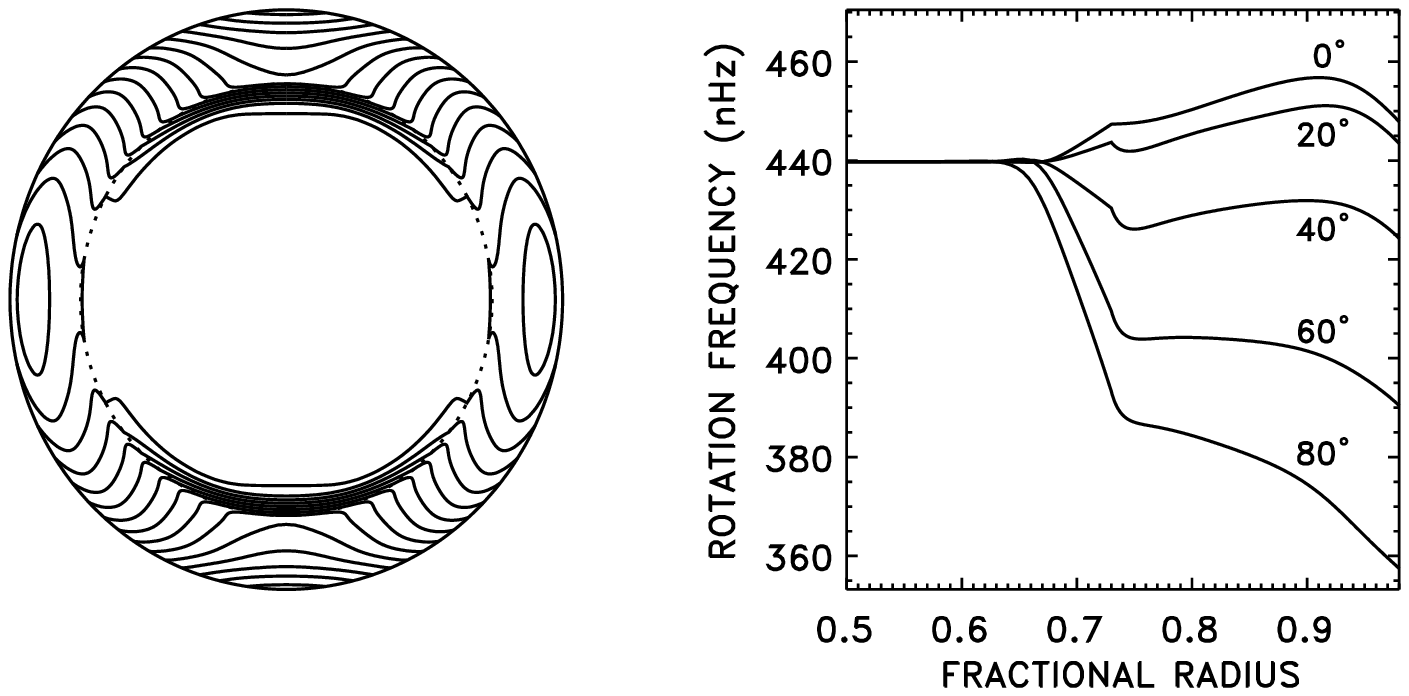}
\end{center}\caption[]{
Rotation law obtained by Kitchatinov \& R\"udiger (2005) taking
the anisotropy of the turbulence near the surface into account.
[Courtesy Kitchatinov \& R\"udiger (2005).]
}\label{kitfig5}\end{figure}

The theory of this negative near-surface shear layer is still a matter
of ongoing research, but it is clear that negative shear would generally
be the result of predominantly vertical turbulent velocities such as
strong downdrafts near the radiating surface.
However, such a layer that is dominated by strong downdrafts
was only thought to be several megameters deep,
and not several tens of megameters.
With an improved theory for the anisotropy of the turbulence especially
near the surface layers, one obtains a clear radial decline of the local
angular velocity near the surface, although still not quite as much
as is observed; see \Fig{kitfig5}.
In any case, these results do at least reproduce
the near-surface shear layer qualitatively correctly.
A proper understanding of this layer is now quite timely in view
of the fact that near-surface shear is likely to contribute to the production
of strong toroidal fields.

\subsection{Magnetic effects}
 
In \Sec{TorsionalOscillations} we mentioned the torsional oscillations,
which is a cyclic modulation of the latitudinal profile of the angular
velocity at the surface of the sun.
Model calculations suggest that these oscillations can well be modeled by
restoring the Lorentz force by adding a term $-\varpi\meanBB\meanB_\phi$
under the divergence in \Eq{AMeqn}.
Unfortunately, given that there is no definitive solar dynamo model,
models for the Sun's torsional oscillations are equally preliminary
and still a matter ongoing research.

In this connection it may be worth noting that there are also magnetic
effects on other properties of the sun, most notably luminosity
variations (by about 0.1\%) and changes of the Sun's quadrupole moment.
The latter does not really seem to be important for the Sun, but in close
binaries this effect leads to measurable changes in the orbital period.

\section{Conclusions}

In the past few decades there have been significant developments
in understanding the physics of the Sun.
Even regarding the radial structure of the Sun, which was thought to be
qualitatively well understood, major revisions have emerged just
recently with the refinement of three-dimensional simulations of solar
granulation.
Such simulations have led to new spectral line fits that imply a
drastically reduced abundance of the heavier elements.
This has consequences for the opacities that affect the deep
parts of the Sun's interior.

There are many aspects of solar physics where a detailed
understanding of the three-dimensional flow pattern of the Sun is
crucial.
It is not surprising that effects involving details of the turbulent
flow field in the solar convection zone, such as the theory of differential
rotation and magnetic field generation, provide other examples where
the three-dimensional dynamics is important.
Fully three-dimensional simulations of solar convection with magnetic
fields produce flow and magnetic field structures in
great detail, but at present they deviate in some important aspects from the Sun
(e.g.\ the fraction of small scale to large scale field is rather large;
and the angular velocity contours are still too strongly aligned with
the rotation axis).
Some tentative explanations are available (magnetic Prandtl number
not small enough in the simulations to reduce or even suppress small
scale dynamo action, and surface conditions not realistic enough
to allow for sufficiently large a baroclinic term).
Future advances in computer technology will bring a steady increase
in numerical resolution.
However, increase of spatial resolution by a factor of two will always be
very difficult when close to the machine capacity.
Substantial progress may rather hinge on new insights that may emerge from a
closer interrelation between local simulations where turbulence is well
resolved and mean field calculations that benefit from input and calibration
of detailed simulations.

\section*{Acknowledgements}

I thank David Moss and an anonymous referee for making useful suggestions
that have helped improving this chapter, and Dr.\ Y. Kamide for his
patience.
I apologize again for not having been able to acknowledge, through
appropriate references, the many achievements that have been reported here.
I have therefore mainly been referencing review material that in turn
quote the relevant original work.

\section*{References}
\begin{list}{}{\leftmargin 1.4em \itemindent -1.4em\listparindent \itemindent
\itemsep 0pt \parsep 1pt}

\item
Bahcall, J. N., Basu, S., \& Serenelli, A. M.\yapjS{2005}{631}{1281}
{1285}{What Is the Neon Abundance of the Sun?}

\item
Benevolenskaya, E. E., Hoeksema, J. T., Kosovichev, A. G., \&
Scherrer, P. H.\yapj{1999}{517}{L163}
{L166}{The interaction of new and old magnetic fluxes at the beginning
of solar cycle 23}

\item
Brandenburg, A.\yapj{2005}{625}{539}
{547}{The case for a distributed solar dynamo shaped by near-surface shear}

\item
Brandenburg, A., \& Tuominen, I.\yjour{1988}{Adv. Space Sci.}{8}{185}
{189}{Variation of magnetic fields and flows during the solar cycle}

\item
Brandenburg, A., \& Subramanian, K.\yjour{2005}{Phys.\ Rep.}{417}{1}
{209}{Astrophysical magnetic fields and nonlinear dynamo theory}

\item
Brandenburg, A., Moss, D., R\"udiger, G., \&
Tuominen, I.\ysphS{1990}{128}{243}{251}
{The nonlinear solar dynamo and differential rotation: A Taylor number puzzle?}

\item
Brandenburg, A., Moss, D., \& Tuominen, I.\yana{1992}{265}{328}
{344}{Stratification and thermodynamics in mean-field dynamos}

\item
Brandenburg, A., Chan, K. L., Nordlund, \AA., \&
Stein, R. F.\yan{2005}{326}{681} 
{692}{Effect of the radiative background flux in convection}

\item
Brun, A. S., Miesch, M. S. \& Toomre, J.\yapj{2004}{614}{1073}
{1098}{Global-scale turbulent convection and magnetic dynamo action in
the solar envelope}

\item
Christensen-Dalsgaard, J., Duvall, T. L., Jr., Gough, D. O.,
Harvey, J. W., \& Rhodes, E. J., Jr.\ynat{1985}{315}{378}
{382}{Speed of sound in the solar interior}

\item
Demarque, P., \& Guenther, D. B.\ypnas{1999}{96}{5356}
{5359}{Helioseismology: Probing the interior of a star}

\item
Deubner, F.-L.\yana{1975}{44}{371}
{379}{Observations of low wavenumber nonradial eigenmodes of Sun}

\item
Dikpati, M., \& Charbonneau, P.\yapj{1999}{518}{508}
{520}{A Babcock-Leighton flux transport dynamo with solar-like
differential rotation}

\item
Duvall, T. L., Jr.\ynat{1982}{300}{242}
{243}{A dispersion law for solar oscillations}

\item
Gailitis, A., Lielausis, O., Platacis, E., et al.\yprl{2001}{86}{3024}
{3027}{Magnetic field saturation in the Riga dynamo experiment}

\item
Gizon, L., \& Birch, A. C.,
``Local Helioseismology,''
{\it Living Rev. Solar Phys.} {\bf 2} (2005),  
\url{http://www.livingreviews.org/lrsp-2005-6}

\item
Gough, D.\ysph{1985}{100}{65}
{99}{Inverting helioseismic data}

\item
Kazantsev, A. P.\yjetp{1968}{26}{1031}
{1034}{Enhancement of a magnetic field by a conducting fluid}

\item
Kippenhahn, R. \& Weigert, A.\ybook{1990}{Stellar structure and evolution}
{Springer: Berlin}

\item
Kitchatinov, L. L. \& R\"udiger, G.\yan{2005}{326}{379}
{385}{Differential rotation and meridional flow in the solar convection 
zone and beneath}

\item
Krause, F., \& R\"adler, K.-H.\ybook{1980}
{Mean-Field Magneto\-hydro\-dy\-na\-mics and Dynamo Theory}
{Pergamon Press, Oxford}

\item
Miesch, M. S., Elliott, J. R., Toomre, J., et al.\yapj{2000}{532}{593}
{615}{Three-dimensional spherical simulations of solar convection.
I. Differential rotation and pattern evolution achieved with laminar and turbulent states}

\item
Marsh, N., \& Svensmark, H.\yssr{2000}{94}{215}
{230}{Cosmic rays, clouds, and climate}

\item
Mihalas, D.\ybook{1978}{Stellar Atmospheres}
{W. H. Freeman: San Francisco}

\item
Moffatt, H. K.\ybook{1978}
{Magnetic field generation in electrically conducting fluids}
{Cambridge University Press, Cambridge}

\item
Ossendrijver, M.\yanar{2003}{11}{287}
{367}{The solar dynamo}

\item
Parker, E. N.\ybook{1979}{Cosmical Magnetic Fields}{Clarendon Press, Oxford}

\item
R\"udiger, G.\ybook{1989}{Differential rotation and stellar convection:
Sun and solar-type stars}{Gordon \& Breach, New York}

\item
R\"udiger, G. \& Brandenburg, A.\yana{1995}{296}{557}
{566}{A solar dynamo in the overshoot layer:
cycle period and butterfly diagram}

\item
R\"udiger, G., \& Hollerbach, R.\ybook{2004}{The magnetic universe}
{Wiley-VCH, Weinheim}

\item
Solanki, S. K., Inhester, B., Sch\"ussler, M.\yrpp{2006}{69}{563}
{668}{The solar magnetic field}

\item
Spruit, H. C.\ysph{1974}{34}{277}
{290}{A model of the solar convection zone}

\item
Stieglitz, R., \& M\"uller, U.\ypf{2001}{13}{561}
{564}{Experimental demonstration of a homogeneous two-scale dynamo}

\item
Stix, M.\yana{1974}{37}{121}
{133}{Comments on the solar dynamo}

\item
Stix, M.\ybook{2002}{The Sun: an introduction}
{Springer-Verlag, Berlin}

\item
Thompson, M. J., Christensen-Dalsgaard, J., Miesch, M. S.,
\& Toomre, J.\yaraa{2003}{41}{599}
{643}{The internal rotation of the Sun}

\item
Ulrich, R. K.\yapj{1970}{162}{993}
{1002}{The five-minute oscillations on the solar surface}

\end{list}
\printindex
\end{document}